\documentclass[final,3p,times,twocolumn,authoryear]{elsarticle}

\usepackage{amssymb}
\usepackage{xspace}
\usepackage{aas_macros}
\usepackage{textcmds}

\journal{New Astronomy}

\newcommand{\kms}{\ensuremath{\mathrm{km\,s}^{-1}}\xspace}
\newcommand{\ergs}{\ensuremath{\mathrm{erg\,s}^{-1}}\xspace}

\newcommand{\gs}{\ensuremath{\mathrm{g\,s}^{-1}}\xspace}

\newcommand{\Msol}{\ensuremath{\mathrm{M}_{\odot}}\xspace}
\newcommand{\Mdot}{\ensuremath{\mathrm{M}_{\odot}}\,\mathrm{yr}^{-1}\xspace}

\begin{document}

\begin{frontmatter}



\title{SS433: a massive X-ray binary at advanced evolutionary stage}

\author[sai]{Anatol Cherepashchuk}
\author[sai]{Konstantin Postnov}
\author[IKI]{Sergey Molkov}
\author[sai]{Eleonora Antokhina}
\author[sai]{Alexander Belinski}

\address[sai]{Sternberg Astronomical Institute, M.V. Lomonosov Moscow State University,
13, Universitetskij pr., 119234, Moscow, Russia}
\address[IKI]{Space Research Institute, 84/32, Profsoyuznaya, 117997, Moscow, Russia}

\begin{abstract}
INTEGRAL IBIS/ISGRI  18-60 keV observations of SS433 performed in 2003-2011 enabled for the first time the hard X-ray phase-resolved orbital and precessional light curves and spectra to be constructed. The spectra can be fitted by a power-law with photon index $\simeq 3.8$ and remain almost constant while the X-ray flux varies by a factor of a few. This suggests that the hard X-ray emission in SS433 is produced not in relativistic jets but in an extended quasi-isothermal hot 'corona' surrounding central parts of a supercritical accretion disc. Regular variations of the hard X-ray flux in SS433  exhibit, on top of the orbital and precessional variability, a nutational variability with a period of $\sim 6.29$~d. For the first time, a joint analysis of the broadband 18-60 keV orbital and precessional light curves was performed in the model assuming a significant Roche lobe overfilling by the optical star, up to its filling the outer Lagrangian surface enabling mass loss through the outer Lagrangian L$_2$ point. From this modeling, the relativistic-to-optical component mass ratio $q=M_x/M_v\gtrsim0.4\div 0.8$ is estimated. An analysis of the observed long-term stability of the orbital period of SS433 with an account of the recent observations of SS433 by the VLTI GRAVITY interferometer enabled an independent mass ratio estimate $q>0.6$. This estimate in combination with the radial velocity semi-amplitude for stationary He II emission, $K_x=168\pm 18\,\kms$ \citep{2004ApJ...615..422H} suggests the optical component mass in SS433 $M_v>12\,\Msol$. Thus, the mass of the relativistic component in SS433 is $M_x>7\,\Msol$, which is close to the mean mass of black holes in X-ray binaries ($\sim 8\,\Msol$). The large binary mass ratio in SS433 allows us to understand why there is no common envelope in this binary at the secondary mass transfer evolutionary stage and the system remains semi-detached \citep{2017MNRAS.471.4256V}. We also discuss unsolved issues and outline prospects for further study of SS433. 

\end{abstract}

\begin{keyword}
binary system \sep evolution \sep jets \sep supercritical accretion disc  \sep precession \sep black hole \sep neutron star \sep accretion \sep microqusar.


\end{keyword}

\end{frontmatter}


\section{Introduction}
\label{Sec:Intro}

SS433 was included by Stephenson and Sanduleak into a Catalog of strong H$_\alpha$ emission sources \citep{1977ApJS...33..459S}. The first optical spectrum of SS433 obtained by \cite{1978Natur.276...44C} revealed the presence of emission lines that could not be identified with known elements. Further spectroscopic studies \cite{1979ApJ...233L..63M,1979ApJ...230L..41M,1980A&A....85...14M,1983A&A...119..153M} discovered moving emission lines of hydrogen and neutral helium shifting with a period of $\sim 164$~d. To explain the moving emission lines, \cite{1979MNRAS.187P..13F} and \cite{1979A&A....76L...3M} suggested a now commonly recognized kinematic model of two oppositely directed relativistic jets moving with a velocity of $v_j\simeq 0.26 c$ ($c$ is the speed of light). This model was supported by the discovery of stationary emission lines Doppler-shifted by a binary orbital motion with period $P_b\simeq 13.1$~d \citep{1980ApJ...235L.131C}.  To explain the precessional variability of SS433, \cite{1980A&A....81L...7V} applied a model of slaved accretion disc formed by precessional motion of the rotational axis of the optical companion \citep{1973SvA....16..756S,1974ApJ...187..575R}. The slaved disc model for SS433 was justified by \cite{1982ApJ...260..780K}. The discovery of optical eclipses in SS433 binary system (V1343 Aql) \citep{1981MNRAS.194..761C} unveiled the nature of this exceptional object as a massive eclipsing close binary system at advanced evolutionary stage with a supercritical accretion disc  \citep{1973A&A....24..337S} around a compact object. This model was further supported by spectroscopic observations by \cite{1981ApJ...251..604C} who measured the radial velocity curve using the stationary HeII 4686 emission line and estimated the mass function of the compact object $f_x(M)\simeq 10\, \Msol$.  

Presently, it is clear that SS433 is a massive Galactic microquasar at supercritical accretion stage with mildly relativistic precessing jets. It is located in the centre of a supernova remnant, plerion W50, at a distance of $d\simeq 5.5$~kpc \citep{1984ARA&A..22..507M,1981MNRAS.194..761C,1989ASPRv...7..185C,2004ASPRv..12....1F}. This object has been intensively studied in the radio, IR, optical and X-ray bands (see \cite{2004ASPRv..12....1F} for a review and early references). The binary system SS433 demonstrates three types of spectral and photometric periodic variability: precessional ($P_{pr}\simeq 162.3$~d), orbital ($P_b\simeq 13.282$ d) and nutational ($P_{nut}\simeq 6.29$ d) ones \citep{1984ARA&A..22..507M,1998ARep...42..336G,1998ARep...42..209G,2008ARep...52..487D} which remains on average stable over about 40 years \citep{2018ARep...62..747C}. The stability of the precessional period supports the slaved disc model for SS433 \citep{2008ARep...52..487D,2018ARep...62..747C}. The stability of the orbital period suggests a large binary mass ratio $q=M_x/M_v\gtrsim 0.6$, where $M_x$ and $M_v$ is the mass of the compact object and the optical star, respectively \citep{2018MNRAS.479.4844C}.

Although SS433 has been studied for about 40 years, the main issue about the nature of the compact object (neutron star or black hole) remains unsolved. The presence of absorption lines in the optical spectrum of SS433 \citep{2002ApJ...578L..67G,2008ApJ...676L..37H} suggests the massive donor star spectral class $\sim$ A7Ib. By assuming that these lines are formed in the optical star photosphere, from the observed Doppler shift of these lines \cite{2008ApJ...676L..37H} estimated the binary mass ratio $q\simeq 0.3\pm 0.11$ and obtained the component masses $M_x=3.3\pm 0.8\,\Msol$, $M_v=12.3\pm 3.3\,\Msol$, which suggests a black hole nature of the compact object. Later careful spectroscopic observations of SS433 by the \textit{Subaru} and \textit{Gemini} telescopes \citep{2010ApJ...709.1374K} led to similar estimates but does not excluded a heavy neutron star. Based on the optical star mass estimate $M_v=8.3-12.5\, \Msol$ obtained for the assumed distance to SS433 and using the binary mass ratio $q=0.15$ as derived from the analysis of 2-10 keV X-ray eclipses \citep{1998IAUS..188..358K,1989A&A...218L..13B,1989PASJ...41..491K}, \cite{2011PZ.....31....5G} concluded that the compact object in SS433 is a neutron star.However, the small $q$ found in these papers in the model of the optical star with sharp limb filling its Roche lobe can be questioned. There are both theoretical \citep{2015MNRAS.449.4415P,2017MNRAS.465.2092P} and observational \citep{2006A&A...460..125F,2009MNRAS.397..479C} evidences that the optical star in SS433 can stably overfill its Roche lobe for a long time so that the 2-10 keV eclipse of the relativistic jets cannot be considered as a purely geometrical screening by the Roche lob-filling donor star. Therefore, the estimate $q\simeq 0.15$ derived from X-ray observations should be considered as a lower limit only \citep{2018MNRAS.479.4844C}.

INTEGRAL observations of SS433 in hard X-rays discovered a hot quasi-isothermal 'corona' above the supercritical accretion disc \citep{2005A&A...437..561C,2007ESASP.622..319C}. These observations also revealed a peculiar variability of the form of and width of the primary X-ray eclipse suggesting its being not a purely geometric. A joint analysis of the eclipse and precessional hard X-ray light curves in the model of Roche-lobe filling optical star yielded the estimate $q\sim 0.3-0.5$. Taking the most likely value $q\simeq 0.3$ and the assumed optical star mass function $f_x(M)=0.268\,\Msol$ \citep{2008ApJ...676L..37H}, the masses of the binary components in SS433 were estimated to be $M_x=5.3\,\Msol$,  $M_v=17.7\,\Msol$ \citep{2013MNRAS.436.2004C}. 

New spectroscopic observations of SS4333 on large telescopes put in doubt the correctness of the optical star spectral class estimate due to strong effects of selective absorption in the powerful stellar wind from the supercritical accretion disc and in a circumbinary gas shell \citep{2008ApJ...678L..47B,2013A&A...556A.149B,2018A&A...619L...4B}. Clearly, the binary mass ratio and the nature of the compact object in SS433 should be precised further from different studies.

Here we highlight the results of the analysis of hard X-ray light curves and spectra of SS433 obtained during decade observations by INTEGRAL.

\section{Hard X-ray spectra at different phases of precessional and orbital variability.}
\label{Sec:Xspectra}

\begin{figure}
	\includegraphics[width=\columnwidth]{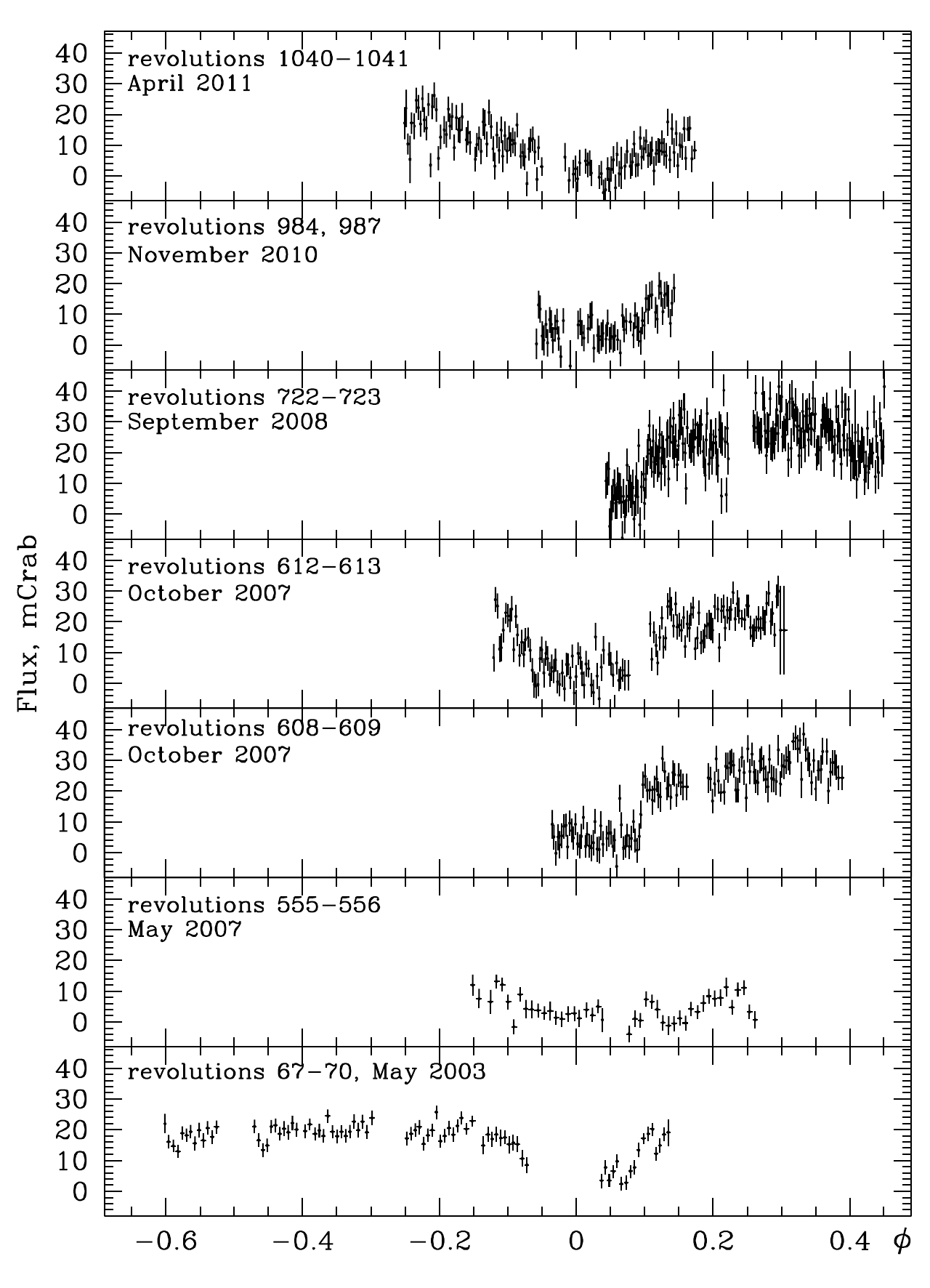}
    \caption{Primary eclipses of SS433 near precession phase zero observed by INTEGRAL. 18-60 keV IBIS/ISGRI fluxes obtained in individual 2 ks INTEGRAL science windows are shown.}
    \label{fig:figure1}
\end{figure}

First observations of SS433 by INTEGRAL discovered for the first time a significant 20-100 keV X-ray flux from SS433 \citep{2003A&A...411L.441C}. Subsequent INTEGRAL pointings enabled us to study the eclipsing and precessional variability of the source in hard X-rays \citep{2013MNRAS.436.2004C}. In total, seven sets of dedicated observations were performed from 2003 to 2011. SS433 was mainly observed close to the primary minimum orbital phases when the accretion disc is seen maximal open to the observer (the precessional phase $\psi_{prec}=0.0$ corresponding to the moment T$_3$ of the maximum separation of moving emission lines, see \cite{1981MNRAS.194..761C}). In some cases \citep{2007ESASP.622..319C} the object was observed at other precessional phases, including those around the moments of the moving emission line cross-over (see Table \ref{T1} and Fig. \ref{fig:figure1} ).

\begin{table}
\caption{Dedicated \textit{INTEGRAL} observations of SS433 primary eclipses at precessional phases with maximum accretion disc  opening}
\label{tabular:IntegralObservations}
\begin{tabular}{llll}
\hline
{\footnotesize Set} & {\footnotesize \textit{INTEGRAL} orbits} &  {\footnotesize Dates} & {\footnotesize Prec. phase $\psi_{pr}$} \\
\hline
I & 67-70 & May 2003 & 0.001-0.060 \\
II & 555-556 & May 2007 & 0.980-0.014 \\
III & 608-609 & October 2007 & 0.956-0.990 \\
IV & 612-613 & October 2007 & 0.030-0.064 \\
V & 722-723 & September 2008 & 0.057-0.091 \\
VI & 984 & November 2010 & 0.870-0.890 \\
 & 987 & November 2010 & 0.930-0.940 \\
VII & 1040-1041 & April 2011 & 0.910-0.950 \\
\hline
\label{T1}
\end{tabular}
\end{table}

\begin{figure}
	\includegraphics[width=\columnwidth]{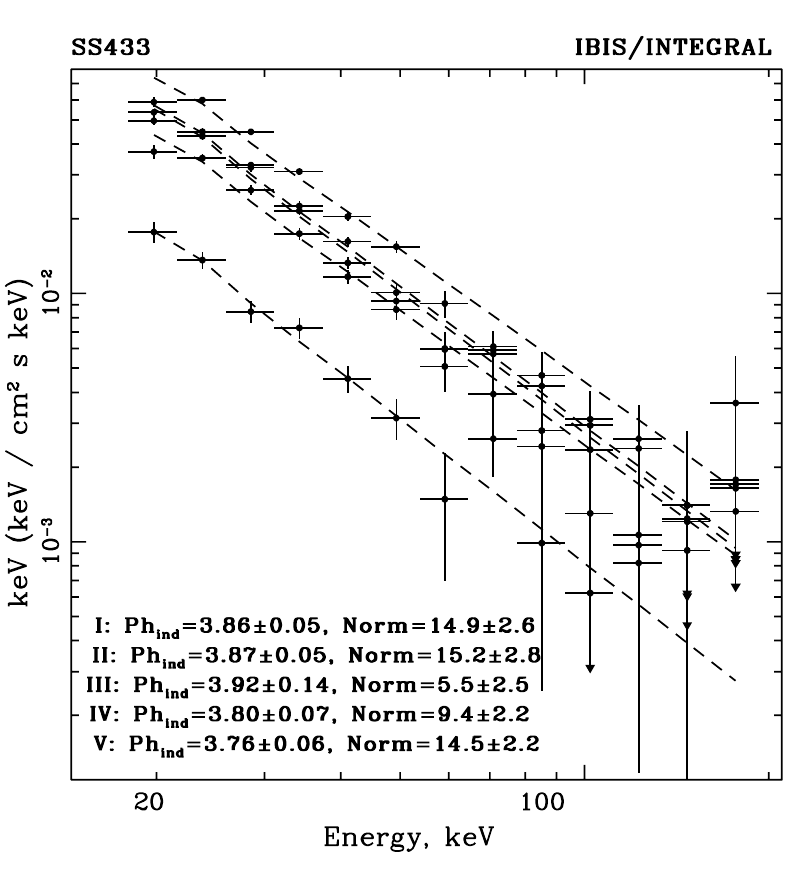}
    \caption{Phase-resolved IBIS/ISGRI spectra of SS433 within selected orbital phase intervals I-V from Fig. 6. 1-$\sigma$ errors are shown.}
    \label{Fig:2}
\end{figure}

\citep{2009MNRAS.397..479C,2013MNRAS.436.2004C} showed that 18-60 keV INTEGRAL spectrum of SS433 keeps its shape at different orbital and precessional phases whereas the X-ray flux from the source varies on average by a factor $\sim 3-4$ (see Fig. \ref{Fig:2}). This suggests that the hard X-ray emission is mainly produced not by relativistic jets\footnote{The relativistic beaming effect due to the precessional and nutational motion of relativistic jets provides minor contribution to the hard X-ray flux variability.}, in which the temperature of thermal plasma rapidly decreases from the jet base, but in a quasi-isothermal extended 'corona' surrounding the supercritical accretion disc. The 18-100 keV X-ray spectrum of the corona can be fit by power-law $dN/dE/dt/dS\sim e^{-\Gamma}$ with the mean photon index $\Gamma\approx 3.8$ for different orbital and precessional phases. A more sophisticated spectral analysis reveals (see Fig. \ref{Fig:2}) a slight spectral hardening at orbital phases $\phi_{orb}=0.25$ (corresponding to interval V ,in Fig. \ref{Fig:2} $\Gamma=3.76\pm0.06$) relative to the phase $\phi_{orb}=0.5$ (corresponding to interval I in Fig. \ref{Fig:2}, $\Gamma=3.86\pm0.05$). In addition, in the middle of the X-ray eclipse ($\phi_{orb}=0$ corresponding to interval III in Fig. \ref{Fig:2}) the spectrum becomes softer $\Gamma=3.92\pm0.14$, although this conclusion is tentative due to large errors. \cite{2013MNRAS.436.2004C} suggested these fine effects with marginal significance might reflect the presence of a hot plasma in a conical funnel around relativistic jets with an opening angle of $\sim 120^o$. Due to the accretion disc nutation (the nutation angle is $\sim 6^o$, \cite{1984ARA&A..22..507M}), the funnel is seen by different angles which can result in variability of the relative fraction of the hard X-ray emission leading to the change in the spectral hardness.  The same effect can be due to partial eclipse of the funnel by the optical star.  

We stress that the reliable detection of the 6.29-d nutational variability with 10-20\% amplitude in SS433 \citep{2013MNRAS.436.2004C} additionally supports the mechanism of hard X-ray formation not in the relativistic jets but in an extended corona above the accretion disc that keep pace with both precessional and nutational motion of the disc. Thus, the X-ray emission region in SS433 includes two components \citep{1992SvA....36..143A,2003A&A...411L.441C,2005A&A...437..561C,2009MNRAS.397..479C,2013MNRAS.436.2004C, 2004ASPRv..12....1F, 2006A&A...460..125F}: a thermal relativistic jet propagating through a conical funnel with opening angle $\sim 120^o$ from the center of the supercritical accretion disc, and a rarefied high-temperature scattering plasma filling this funnel  (the corona). \cite{2009MNRAS.394.1674K} showed that the observed 3-100 keV emission of SS433 can be described by a two-component model including thermal radiation from the jets (which is responsible for soft X-ray emission)  and Comptonized radiation from the hot corona forming hard X-ray spectrum. The temperature of the corona was found to be $T_{cor}\approx 20$ keV, the Thomson optical depth $\tau_T\approx 0.2$, the mass-loss rate through the jets $\dot M_j\simeq 3\times 10^{19} \gs \,(3\times 10^{-7} \Mdot)$. The electron number density in the corona was assumed to be $n_e\simeq 5\times 10^{12}$ cm$^{-3}$, corresponding to wind outflow with the velocity $\sim 3000\, \kms$ from the supercritical accretion disc with a rate of $\sim 10^{-4}\Mdot$ at a distance of $\sim 10^{12}$~cm from the disc center where the optically thick photosphere is formed in the wind \citep{2004ASPRv..12....1F}.  The disc photosphere size was independently estimated from the analysis of fast optical variability of SS433 \citep{2011AstL...37..100B}. 

This model of the hard X-ray source was used to describe the orbital and precessional X-ray light curves of SS433 observed by INTEGRAL.  

\section{Precessional variability}
\label{Sec:Xprec}

Precessional light curves were constructed using all dedicated pointings and publically available off-eclipse INTEGRAL observations of SS433 when the source fell within the field of view of the INTEGRAL instruments ($\lesssim 13^o$) with a total exposure time of $\sim 8.5 $ Ms. To analyze the precessional variability, each off-eclipse pointing with an exposure time of $\sim 2-5$ ks was assigned with the corresponding precession phase. The orbital and precessional phases were calculated using the ephemeris from \cite{2004ASPRv..12....1F}:

1. The orbital eclipse minimum corresponding to the orbital phase $\phi_{orb}=0$ when the optical star is in front of the accretion disc: JD$_{min}$(hel)=2450023.62+13$^d$.08211$\cdot E$

2. The zero precession phase corresponding to the T$_3$ moment of the maximum separation of the moving emission lines ($\psi_{prec}=0$, the disc is maximum open for the observer) 
JD$_{max}$(hel)=2443507.47+162$^d$.375$\cdot E_1$.

With an account of the distance to SS433 5.5 kpc \citep{2004ASPRv..12....1F}, the observed 18-60 keV luminosity at maximum is $3\times 10^{35} \ergs$.

\begin{figure}
	\includegraphics[width=\columnwidth]{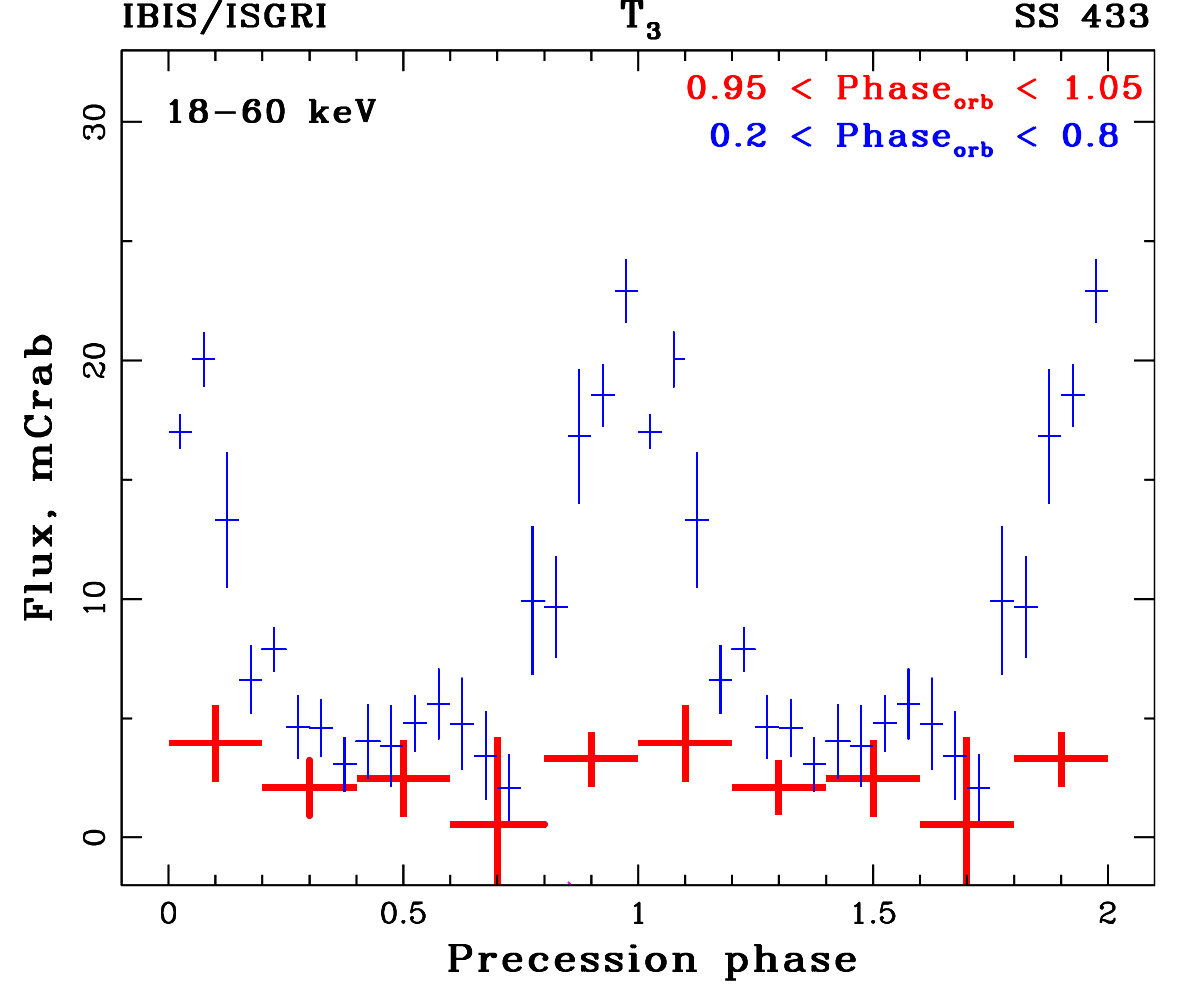}
    \caption{Off-eclipse and mid-eclipse precessional 18-60 keV light curves (upper thin crosses and thick red crosses, respectively).}
    \label{Fig:3prec2}
\end{figure}

Precessional off-eclipse ($0.2<\phi_{orb}<0.8$) and mid-eclipse ($0.95<\phi_{orb}<1.05$) light curves of SS433 are shown in Fig. \ref{Fig:3prec2}. Clearly, the off-eclipse precessional variability is very reliable and remains stable during many precessional periods. The maximum flux from SS433 ($\sim 20$ mCrab) is observed at the the precessional phase $\psi_{prec}\simeq 0$, the minimum flux (the upper limit of minimum flux is $\sim 2-3$ mCrab) is measured at $\psi_{prec}^{1}\simeq 0.34$ and $\psi_{prec}^{2}\simeq 0.66$, roughly corresponding to the cross-over moments when the disc is seen edge-on. A secondary maximum with a flux of $\sim 5-6$ mCrab is seen at precessional phases  $\psi_{prec}\simeq 0.5-0.6$.  Thus, the flux in the primary and secondary maximum of the precessional light curve changes by a factor of $\sim 7$ and $\sim 2$, respectively. In the X-ray eclipse minimum, no significant precessional variability  was detected (Fig. \ref{Fig:3prec2}).

\begin{figure}
	\includegraphics[width=\columnwidth]{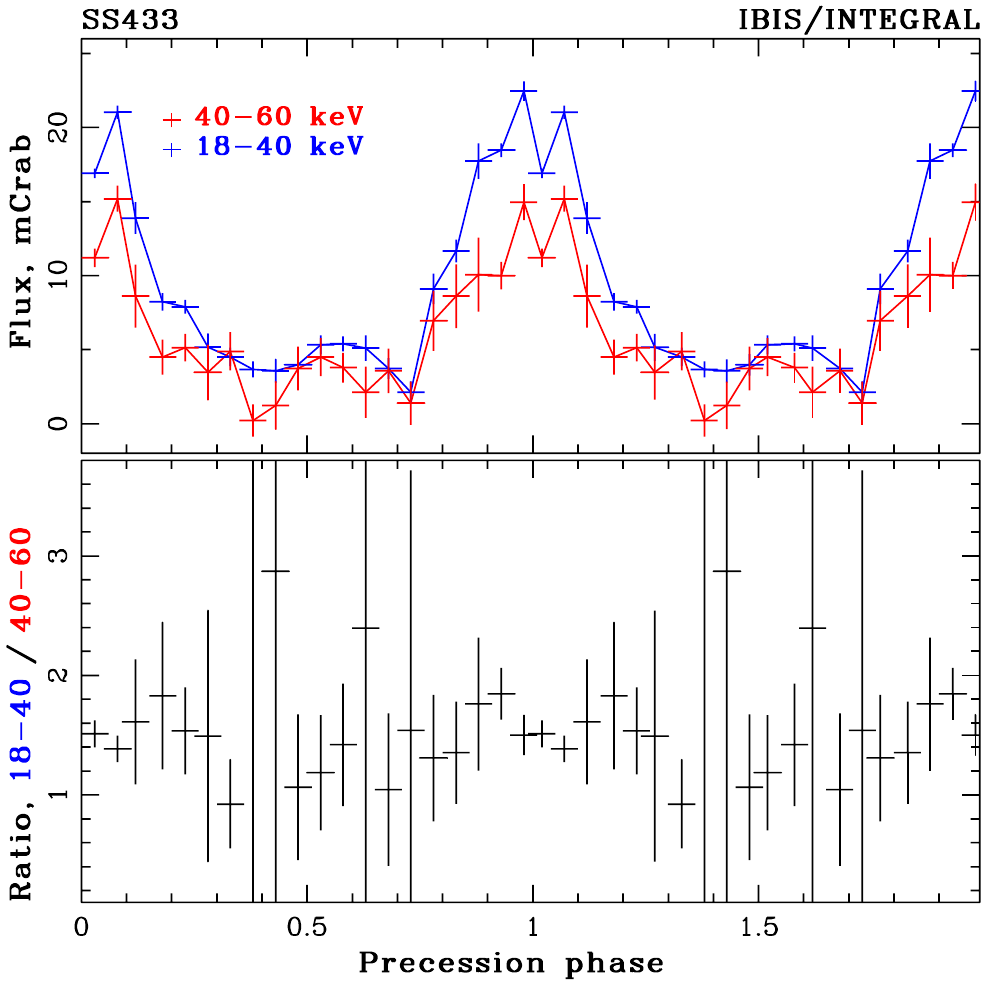}
    \caption{Precessional 18-40/40-60 keV light curves (upper panel) and the spectral hardness ratio (bottom panel). Orbital eclipses are excluded.}
    \label{Fig:4Xprec}
\end{figure}
To analyze the spectral dependence of the precessional variability, the broadband 18-60 keV interval was split into two bands: 18-40 keV and 40-60 keV. Fig. \ref{Fig:4Xprec} shows the corresponding precessional off-eclipse light curves and the dependence of the spectral hardness on the precessional phase. No clear dependence of the X-ray hardness on the precessional phase is seen. Relative precessional variability amplitude is higher in the 40-60 keV band than in 18-40 keV band. This effect, as well as the monotonic increase of the X-ray eclipse depth with energy, was first noted in \citep{2005A&A...437..561C}. The increase in the precessional and orbital variability amplitude with photon energy reflects the decrease in the size of X-ray emitting region, the 40-60 keV emitting region being most compact and without any contribution of thermal radiation from relativistic jets. The hard X-ray precessional variability enables us to constrain the characteristic 'vertical' size of the emission region (normal to the accretion disc plane) while the orbital variability constrains the 'horizontal' size of the emission region. Thus, such a two-dimensional 'tomography' of the disc corona by the precessional and orbital light curves enabled us to obtain independent constraints on the binary system parameters including the component mass ratio in SS433 \citep{2009MNRAS.397..479C,2013MNRAS.436.2004C}.

\section{Orbital light curves}
\label{Sec:Xorb}

\begin{figure}
	\includegraphics[width=\columnwidth]{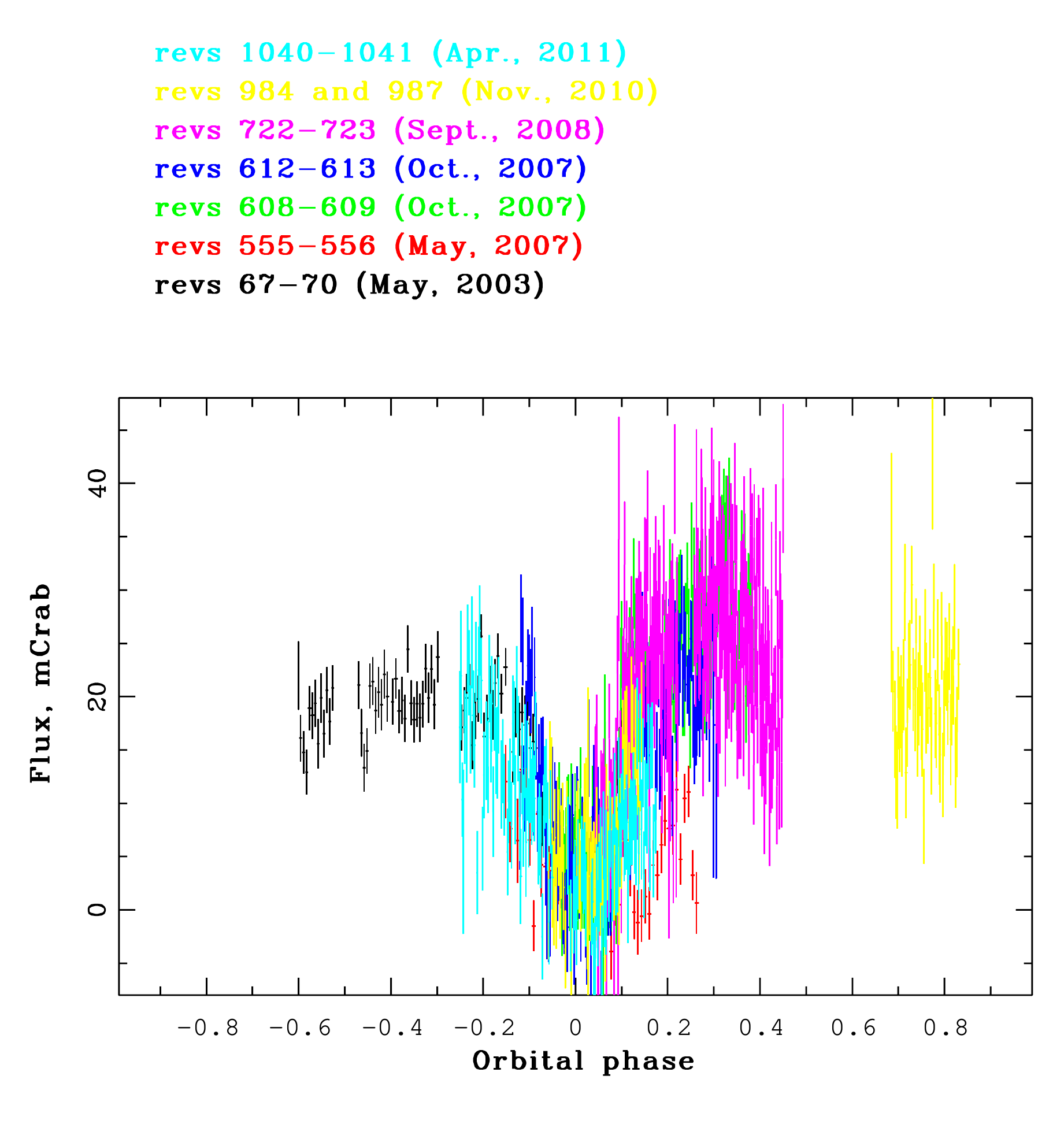}
    \caption{Composite IBIS/ISGRI 18-60 keV X-ray eclipse light curve around precessional phase zero ($0.08 <  \psi_{prec} < 1.2$) for observations from Table 1.}
    \label{Fig:5orb}
\end{figure}

To construct the mean orbital light curve of SS433 we have used dedicated observations carried out mainly around the precessional phase $\psi_{prec}=0$ when the accretion disc is maximum open. The hard X-ray light curve convolved with the orbital period is shown in Fig. \ref{Fig:5orb}. The IBIS/ISGRI INTEGRAL data were analyzed using the software elaborated at IKI by E.M. Churazov (see \cite{2004AstL...30..534M} and references therein).

Individual observations of SS433 show that the orbital light curve of the source is strongly variable. For example, in May 2007 (see Fig. \ref{fig:figure1}) the X-ray eclipse width is much wider than on average, whereas in October 2007 the X-ray width was observed to have average width.  Strong variations of the X-ray eclipse width may suggest that the egress from the eclipse is distorted due to absorption of the X-ray emission by strongly inhomogeneous gas flows from the donor star and by the powerful asymmetric wind from the precessing supercritical accretion disc. Additional absorption can be also produced by a dense region of interacting stellar winds from the donor star and the supercritical accretion disc \citep{1995Ap&SS.229...33C}. Thus, following \cite{2009MNRAS.397..479C}, in our modelling of the X-ray eclipse to infer the SS433 binary parameters, we will use the most stable ingress part of the X-ray orbital light curve that apparently is less subjected to the absorption effects discussed above. Besides, the nutational variability of SS433 with a period of $\sim 6.29$~d and an amplitude of $\sim 20\%$ of the non-eclipsed flux discovered in 18-60 keV \citep{2013MNRAS.436.2004C} can additionally distort the eclipse form. The nutational sine-like variability is seen in Fig. \ref{Fig:5orb} as a continuation of the egress part of the orbital light curve. At the precessional phase close to T$_3$ moments considered here, the ingress part of the light curve is observed to be less subjected to the nutational variability.

\begin{figure}
	\includegraphics[width=\columnwidth]{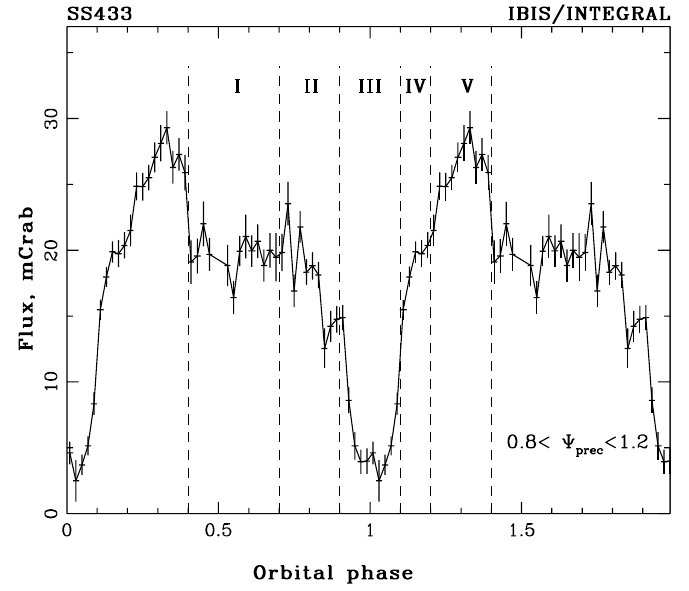}
    \caption{Binned 18-60 keV IBIS/ISGRI  light curve ($\Delta \phi_{orb}$ = 0.02) with orbital phase intervals I-V selected for spectral analysis (vertical dashed lines).}
    \label{Fig:6spint}
\end{figure}

When constructing the mean light curve we have averaged the X-ray fluxes inside phase intervals $\Delta\phi_{orb}=0.02$, which increased the signal-to-noise ratio. Here we omitted the May 2007 observations in which an anomalously wide X-ray eclipse was observed. The mean orbital 18-60 keV light curve is presented in Fig. \ref{Fig:6spint} (bars show $3\sigma$-errors). 
The mean X-ray flux in the X-ray eclipse minimum can be non-zero (the upper limit about $2$  mCrab), possibly due to the contribution of the radiation scattered in the disc wind. 

\label{Fig:5prec2}

Phase intervals I-V in which spectral analysis was performed are noted (see Fig. \ref{Fig:2}). As noted above, spectra calculated for phases I-V do not differ significantly although some spectral hardening with increasing X-ray flux can be noticed. More reliable spectral changes with orbital phase could probe the structure of the hard X-ray emitting region (see, for example, the discussion in \cite{2006A&A...460..125F} and \cite{2009MNRAS.394.1674K}). 

\begin{figure}
	\includegraphics[width=\columnwidth]{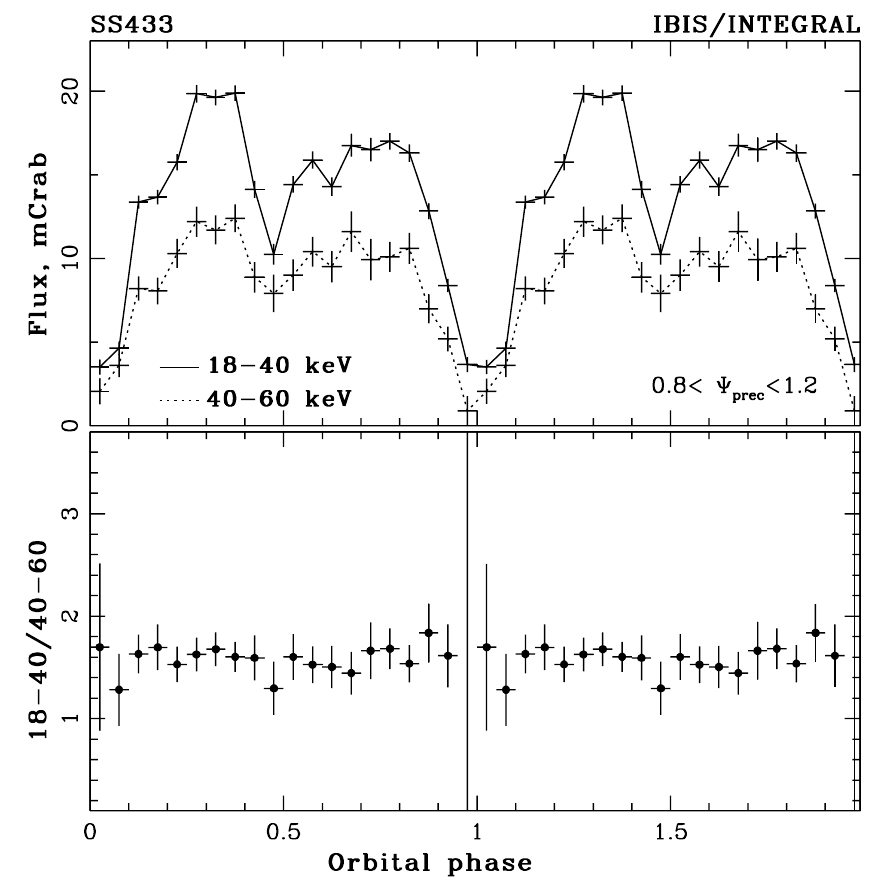}
    \caption{IBIS/ISGRI 18-40 and 40-60 keV orbital light curves (the upper and lower curves, respectively, on the upper panel) with the spectral hardness ratio (bottom panel). }
    \label{Fig:7orb2}
\end{figure}
\label{Fig:6Xorb}
\label{Fig:8orb2bands}

The mean orbital light curves of SS433 in 18-40 keV and 40-60 keV bands are presented in Fig. \ref{Fig:7orb2}. Here also we show the dependence of the X-ray spectral hardness on the orbital phase, which is not significant despite the flux change by an order of magnitude. In these light curves, the nutational variability at phases 0.25-0.75 is clearly visible. Note that thermal radiation from relativistic jets virtually  does not contribute to the orbital 40-60 keV X-ray light curve. 

\section{Geometrical model of SS433}
\label{Sec:mathmodel}

INTEGRAL observations of SS433 can probe the parameters of this unique binary system using three independent light curves. 

1. Precessional off-eclipse 18-60 keV light curve and corresponding 18-40 keV and 40-60 keV light curves (see Figs. \ref{Fig:3prec2}, \ref{Fig:4Xprec}). 

2. Precessional light curve in the middle of eclipse by the donor star (Fig. \ref{Fig:3prec2}). 

3. Orbital 18-60 keV light curve (Fig. \ref{Fig:6spint}) and 18-40 keV and 40-60 keV light curves (Fig. \ref{Fig:7orb2}).

To model the observed X-ray light curves, we have used a geometrical model described in detail in \cite{1992SvA....36..143A,2005A&A...437..561C,2009MNRAS.397..479C,2013MNRAS.436.2004C}.  We consider a close binary system consisting of an opaque and X-ray dark donor star filling or overfilling its Roche lobe and a relativistic object surrounded by an optically and geometrically thick supercitical 'accretion disc' inclined to the orbital plane by the angle $\theta\approx20^o$. The 'accretion disc' includes a proper disc and an extended photosphere  formed by outflowing wind. The disc precesses with a period of 162.3  d. 

\begin{figure}
	\includegraphics[width=\columnwidth]{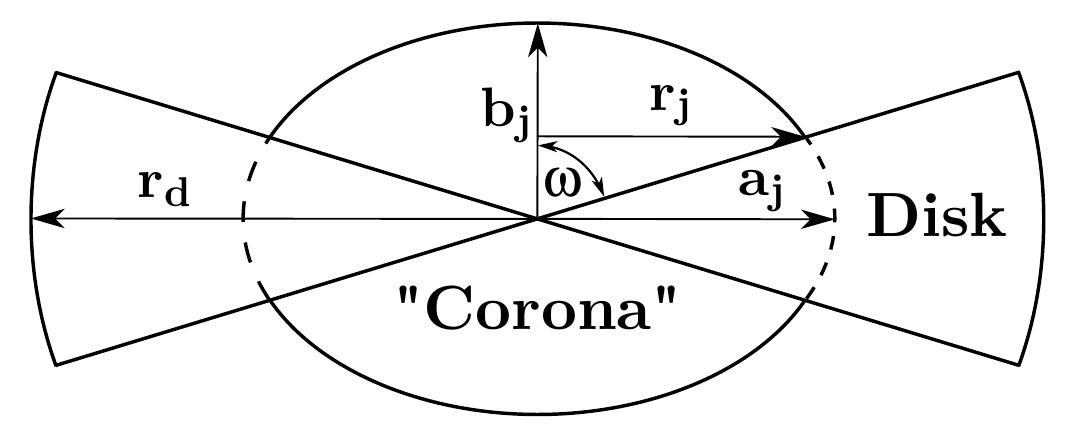}
    \caption{Geometrical model of the accretion disc  and its “corona”.}
    \label{Fig:8}
\end{figure}

\label{Fig:9model}
The opaque and X-ray dark disc body (see Fig. \ref{Fig:8}) is characterized by radius $r_d$ and half-opening angle $\omega$. The central relativistic object is surrounded by an optically thin X-ray emitting spheroidal 'corona' with semi-axes $a_j$ and $b_j$. The visible radius of the 'corona' $r_j$ is determined by the parameter combination $a_j,b_j,\omega$. Depending on the ratio $b_j/a_j$, this structure can be viewed as either oblate 'corona' ($b_j/a_j<1$) or (non-relativistic) prolate 'jet' ($b_j/a_j>1$). The parameters $r_d,a_j,b_j$ and $r_j$ are in units of the relative orbit radius $a$. The size of the donor star is limited either by the inner Lagrangian surface (the Roche lobe)  (passing through the inner Lagrangian point L$_1$) or outer Lagrangian surface (passing through the Lagrangian point L$_2$).  These sizes are uniquely determined by the component mass ratio $q=M_x/M_v$, where $M_x$ and $M_v$ is the mass of the relativistic object and the donor star, respectively. The binary orbit inclination $i\simeq79^o$ and disc tilt $\theta$ are fixed from the analysis of the moving emission line kinematics. 

In our model, only the corona radiates in hard X-rays (18-60 keV) while the donor star and the disc are opaque and screen the corona during the orbital and precessional motions. 
In the precessional motion, the disc inclination angle to the line of sight changes and the inner corona is screened differently by the outer parts of the accretion disc. Thus, the modelling of the precessional light curve enable us to obtain a 'vertical scan' of the corona probing the parameters $b_j$ and $\omega$. On the other hand, the modelling of the orbital light curve yields a 'horizontal scan' of the corona (in the accretion disc plane) enabling the estimate of the parameters $q, r_d, \omega$ and $a_j, b_j$. A joint analysis of the precessional and eclipsing variabilities hence enable us to probe the spatial structure of the corona in the central parts of the disc where the hard X-ray emission is formed. 

The orientation of the system relative to the observer is described by the orbital phase angle (the orbit is assumed to be circular), as well as by the orbital inclination angle $i\simeq79^o$, the accretion disc inclination angle to the orbital plane $\theta\simeq20^o$ and the precessional phase $\psi_{prec}$.  Here the zero precessional phase $\psi_{prec}=0$ corresponds to the moment of the maximum disc opening to the observer (the moment $T_3$ of the maximum separation of the moving emission lines, see \cite{1981MNRAS.194..761C}). At precessional phases $\psi_{prec}=0.34$ and 0.66 the disc is observed edge-on (the cross-over moments $T_1$ and $T_2$ for the moving emission lines, respectively). 

\subsection{Justification of an extended emission region (corona) in SS433}

There are observationally motivated physical grounds for the existence of a hot 'corona' above the accretion disc center. 

1. The precessional and orbital independence (within the measurement errors) of hard X-ray 18-60 keV spectra of SS433 in spite of an order or magnitude changing in X-ray flux. In the 2-10 keV X-ray band, the thermal X-ray emission from relativistic jets dominates. The temperature of plasma strongly decreases along the jets, and significant softening of X-ray spectrum is observed during the eclipse (see, for example, \citep{2006A&A...460..125F}). 

\begin{figure}
	\includegraphics[width=\columnwidth]{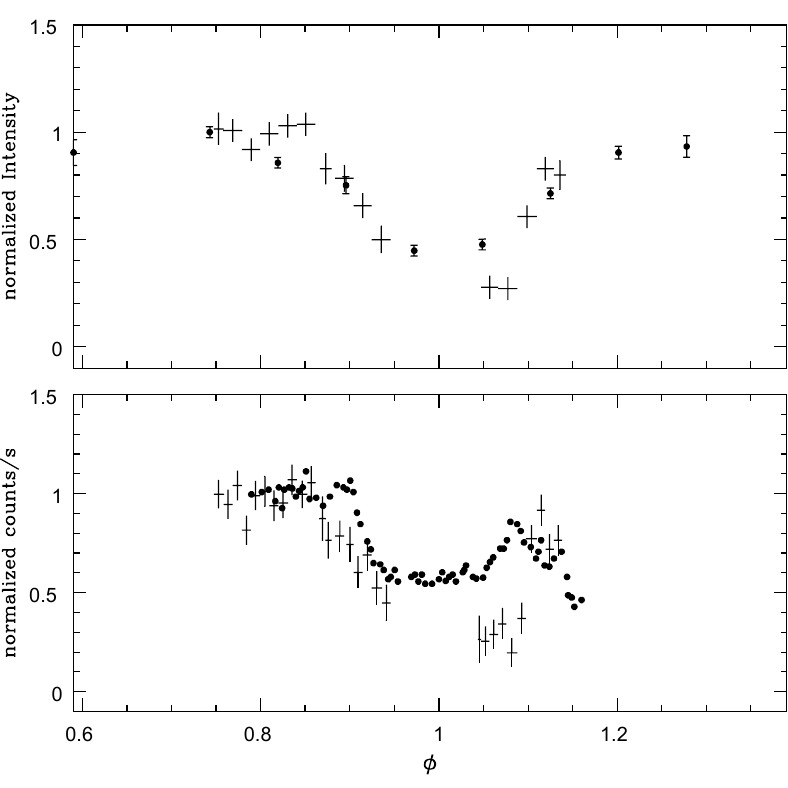}
    \caption{The primary X-ray eclipse of SS433 in the 18-60 keV energy band. IBIS/ISGRI data (May 2003). Upper panel: X-ray light curve averaged over 20000 s (10 INTEGRAL science windows, SCW) superimposed on the simultaneously obtained V optical photometric lightcurve (Crimea, SAO). Bottom panel: the same hard X-ray eclipse light curve averaged over 10000 s (5 SCW) super-imposed on the Ginga 4.6-27 keV eclipse (filled circles; from \cite{1989PASJ...41..491K}, \cite{1995A&A...297..451Y}) taken at about the same precession phase. The INTEGRAL data is the same in both panels, but the averaging is different.}
    \label{Fig:9}
\end{figure}

\label{Fig:10xecl}
2. The 18-60 keV X-ray eclipse by the donor star is wider than 4.6-27 keV eclipse \citep{1989PASJ...41..491K, 1995A&A...297..451Y} (see Fig. \ref{Fig:9}. The increase in the X-ray eclipse width with photon energy is unique property of SS433 suggesting that the characteristic size of the emitting region in harder X-ray band increases and not decreases, as observed in ordinary (not superaccreting) X-ray binaries. 

3. The discovery of hard X-ray nutational variability in SS433 \citep{2013MNRAS.436.2004C}  also suggests that the hard X-ray emission is formed not in the relativistic jets but in an extended hot corona that precesses and nutates coherently with the disc. The hard X-ray flux variability due to relativistic beaming in SS433 with bulk velocity $v\approx 0.26c$ can be $\sim 15\%$ and up to 100\% for nutational and precessional motion of one jet, respectively. However, as in the 18-60 keV band the thermal emission of jets base with $kT\approx 20$ keV \citep{2009MNRAS.394.1674K} should show exponential cut-off, the contribution of the jet thermal emission is insignificant compared to the corona with $L_x\sim 5\times 10^{35}\ergs$. Therefore, we can neglect the relativistic beaming effects and consider the observed hard X-ray precessional and nutational variability as being due to geometrical screening of the corona by the accretion disc body. 

The hot corona can be produced by the collisions of the supercritical accretion disc wind inhomogeneities with the relativistic jets \citep{2006MNRAS.370..399B}. A disc corona can also be formed by convective motions of acoustic waves in the disc \citep{1976SvAL....2..191B,1977A&A....59..111B} or due to magnetic field dissipation \citep{1979ApJ...229..318G}. 
A detailed modelling of the broadband 3-90 keV spectrum of SS433 by emission from accretion disc, relativistic jets an hot corona was carried out by \cite{2009MNRAS.394.1674K}. 

The presence of the hot extended region emitting in hard X-ray in the center of the accretion disc was independently inferred from excess of hard X-ray emission found in \textit{Suzaku} observations of SS433  \citep{2010PASJ...62..323K}. 

Recently, \cite{2018arXiv181010518M} reported NuSTAR 3-60 keV observations of SS433 and proposed a model of the supercritical accretion disc with the central cone-like 'walls' formed by the radial wind outflow. In this model, the hard X-ray emission is not relativistically beamed and is formed in the central extended region of the disc by scattering of accretion-generated hard X-ray photons on the 'walls'. Although the physical generation mechanism of hard X-ray radiation with $kT\sim 20-40 $ keV in this case differs from that explaining the presence of the hot extended 'corona' used in our model, this does not affect our geometrical interpretation of the precessional and orbital variability of SS433, which is based upon  screening of an extended emission region by precessing disc and opaque optical star.

\subsection{Justification of stable Roche lobe overfilling by the donor star in SS433}

In \citep{2013MNRAS.436.2004C}, a joint interpretation of the orbital and precessional hard X-ray variability of SS433 was performed using the geometrical model of accretion disc with hot extended corona and the likely estimate of the component mass ratio in SS433 was obtained, $q=0.3\div 0.5$. The optical star was assumed to fill its Roche lobe and to transfer mass onto the relativistic object through the inner Lagrangian point L$_1$. 

Recently, detailed hydrodynamical simulations through the inner Lagrangian point L$_1$ appeared \citep{2015MNRAS.449.4415P,2017MNRAS.465.2092P} to show that when treating the mass transfer through a de Laval nozzle at L$_1$, the mass transfer rate is restricted by gasdynamic and thermodynamic effects, and the donor star can stably overfill its inner Roche lobe and even provide mass transfer through the outer L$_2$ point. 

The uniqueness of SS433 as a massive X-ray binary system  at the final stage of the secondary mass transfer is that despite the donor star overfills its Roche lobe, no common envelope stage ensures (as predicted by the theory of close binary system evolution, e.g. \cite{MT1988}). The system remains being semi-detached. In this case, the mass and angular momentum outflow from the system is
mediated by the formation of a supercritical accretion disc around the relativistic object with a powerful wind outflow ($v_w\simeq 2000\, \kms,\, \dot M_w\simeq 10^{-4}\Mdot$) and relativistic jets ($v_j\approx 80000\,\kms, \,\dot M_j\simeq 10^{-6}\div 10^{-7}\Mdot$).

The peculiar properties of SS433 were recently analyzed by \cite{2017MNRAS.471.4256V}. The authors concluded that if the donor star with radiative envelope in a massive X-ray binary system starts filling its inner Roche lobe  and the binary mass ratio is $q=M_x/M_v\gtrsim 0.29$, the formation of the common envelope can be avoided. The system evolves as a semi-detached binary with fast but stable overfilling of its Roche lobe. In this case the matter transferred from the donor star through the inner Lagrangian point L$_1$ is expelled from the vicinity of the relativistic object (the so-called isotropic re-emission mode, or 'SS433-like' mode of mass loss).  Oppositely, if $q\lesssim 0.29$,  the massive X-ray binary inevitably plunges into a common envelope during the secondary mass transfer episode. In this case, depending on the initial orbital angular momentum, either a very close WR+BH system (like Cyg X-3) or a Thorne-Zytkow object \citep{1977ApJ...212..832T} is formed. The authors stress that if the relativistic object is a neutron star, the binary mass ratio is likely to be much smaller than 0.29, and the binary should evolve via the common envelope stage with the most likely formation of a Thorne-Zytkow object (unless the orbital period at the beginning of the mass transfer is longer than about 100 d).     
Recently, a red giant candidate for Thorne-Zytkow objects was reported (2014) based on chemical abundance anomalies \cite{2014MNRAS.443L..94L},

Thus, both gasdynamic and evolutionary considerations do not exclude the possibility that the donor star in SS433 can significantly overfill its Roche lobe. Therefore here we performed the modelling of precessional and orbital X-ray light curves of SS433 by assuming that the donor star in SS433 fills its outer critical Roche lobe passing through L$_2$ point. Strictly speaking, when the mass transfer from the optical star occurs through the L$_1$ point and the star fills its outer Roche lobe, the shape of the star can deviate from equipotentials calculated in the classical Roche model.  Moreover, the precessional motion of the donor as suggested by the slaved disc model in SS433 might additionally disturb the shape of the outer parts of the star thus promoting the inner Roche lobe overflow.  The star should 'plump' predominantly in the direction normal to the line connecting centers of the binary components by keeping a relatively narrow 'neck' near the L$_1$ point. However, as we consider only phases of eclipse of the accretion disc by the optical star where the L$_1$ point is screened by star's body, in the first approximation we can treat the shape of the eclipsing star as being limited by the outer critical Roche lobe passing through the L$_2$ point. In this case, complicated gasdynamic effects near the L$_1$ point does not affect the results of interpretation of the eclipse of the central parts of accretion disc with hot corona by the donor star. 

\section{Parameters of SS433 from the analysis of orbital and precessional light curves obtained by INTEGRAL}
\label{Sec:7modelparam}

\subsection{Model with exact Roche lobe filling by the donor star}
\label{Susec:exactRoche}

\begin{figure}
	\includegraphics[width=\columnwidth]{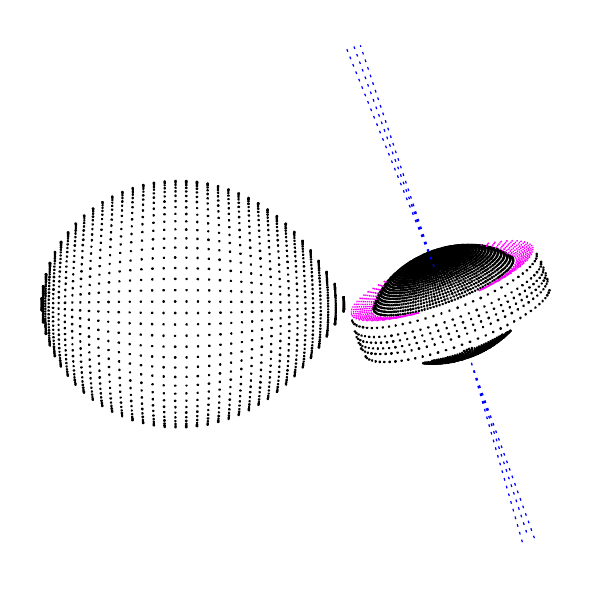}
    \caption{Model view of SS433 for binary mass ratio $q = 0.3$ is shown at the disc  precession angle 20 degrees. Thin jets normal to the disc  plane are shown but their contribution to hard X-ray emission is ignored.}
    \label{Fig:10}
\end{figure}

Here we briefly describe the results of the joint modelling of the precessional and orbital light curves of SS433 carried out in  \citep{2013MNRAS.436.2004C} by the model of SS433 including a supercritical accretion disc with hot corona and a donor star exactly filling its inner Roche lobe (see Fig.  \ref{Fig:10}). The binary inclination angle was fixed at $78^o.8$ and the accretion disc tilt to the orbital plane was fixed at $\theta =20^o.3$ from independent measurements. The fitting parameters were $q, r_d, \omega, a_j, b_j$ (see Fig. \ref{Fig:8}). The weighted sum of deviations between the observed and model light curves (precessional and orbital) was used as the residual. The minimization of the residual functional was performed using the simplex algorithm (Nelder and Mead's method) \citep{10.1093/comjnl/7.4.308,1987ApJ...313..346K}.

\begin{figure}
\center{\includegraphics[width=0.75\columnwidth]{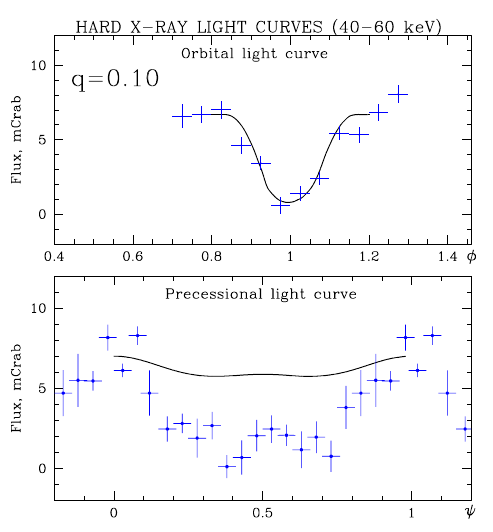} \\(a)	\\}
\center{\includegraphics[width=0.75\columnwidth]{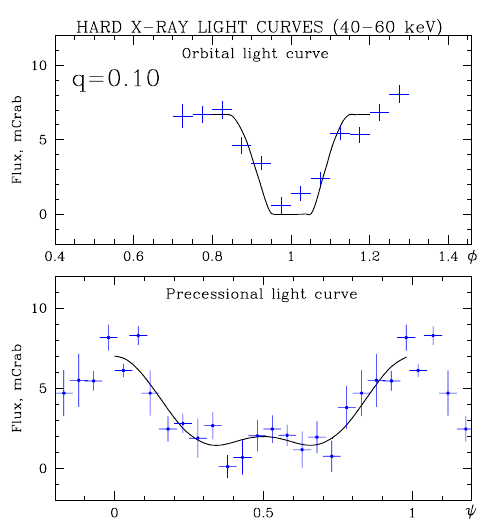} \\(b)	\\}
\center{\includegraphics[width=0.75\columnwidth]{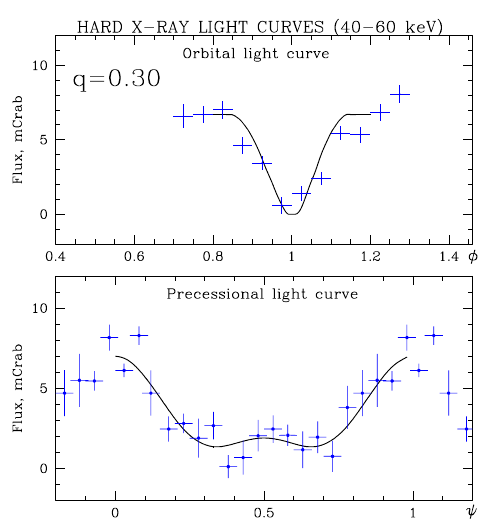} \\(c)	\\}
    \caption{Joint analysis of orbital (upper plots) and precessional (lower plots) 40-60 keV IBIS/ISGRI light curves of SS433. Panel (a): $q = 0.1$, ’long jet’ corona ($a_j = 0.25$, $b_j = 0.55$, $\omega = 80^o$); only orbital light curve can be reproduced. Panel (b): $q = 0.1$, ’short jet’ corona ($a_j = 0.25$, $b_j = 0.1$, $\omega = 80^o$). Both orbital and precessional light curvs can be fitted, but the total eclipse (plateau at zero flux) appears (unobserved). Panel (c): $q = 0.3$, ’short jet’ corona ($a_j = 0.35$, $b_j = 0.13$, $\omega = 80^o$); both orbital and precessional light curves are well reproduced.}
    \label{Fig:11}
\end{figure}

The results of the joint analysis of the orbital and precessional 40-60 keV  light curves to which thermal X-ray emission from relativistic jets can be neglected are shown in Fig. \ref{Fig:11} a,b,c. These results fully confirm our earlier analysis \citep{2009MNRAS.397..479C}.

Fig. \ref{Fig:11} suggest that although the eclipsing light curve alone can be well fitted assuming a low binary mass ratio $q\sim 0.1$ in the model with 'long' thin X-ray jet ($b_j\gtrsim0.5$), the corresponding precessional light curve cannot be reproduced for $q\lesssim 0.25$ (see Fig. \ref{Fig:11} a) because the precessing opaque accretion disc screens only small part of the long thin X-ray jet. On the other hands, a model with 'short' X-ray jet, while enabling good description of the precessional light curve for $q\sim 0.1$, is unable to fit the orbital eclipse. In the latter case, the 'short' thin jet is fully eclipsed by the donor star resulting in a noticeable plateau at the eclipse minimum, contrary to what is actually observed (Fig. \ref{Fig:11} b). For $q=0.3$ both orbital and precessional light curves can be well fitted (Fig. \ref{Fig:11} c). 

Fig. \ref{Fig:deviations} (upper panel) shows the $\chi^2$ residuals for the orbital 18-60 keV light curve minimized for all parameters but $q$, as a function of the mass ratio $q$. The model parameters provide good fitting to the precessional light curve (see Fig. \ref{Fig:11} b, c, bottom panel).  Fig. \ref{Fig:deviations} suggests that the best joint fit of the orbital and precessional light curves in the model of the precise filling of the Roche lobe by the optical donor star is obtained for $0.3\lesssim q\lesssim 0.5$. 

Note that the estimate $0.3\lesssim q\lesssim 0.5$ derived from the modeling of hard X-ray light curves
is consistent with that obtained earlier from the modeling of optical orbital and precessional light curves of SS433 \citep{1987SvA....31..295A}.

\subsection{Model with the donor star inner Roche lobe overflow}

\begin{figure}
	\includegraphics[width=\columnwidth]{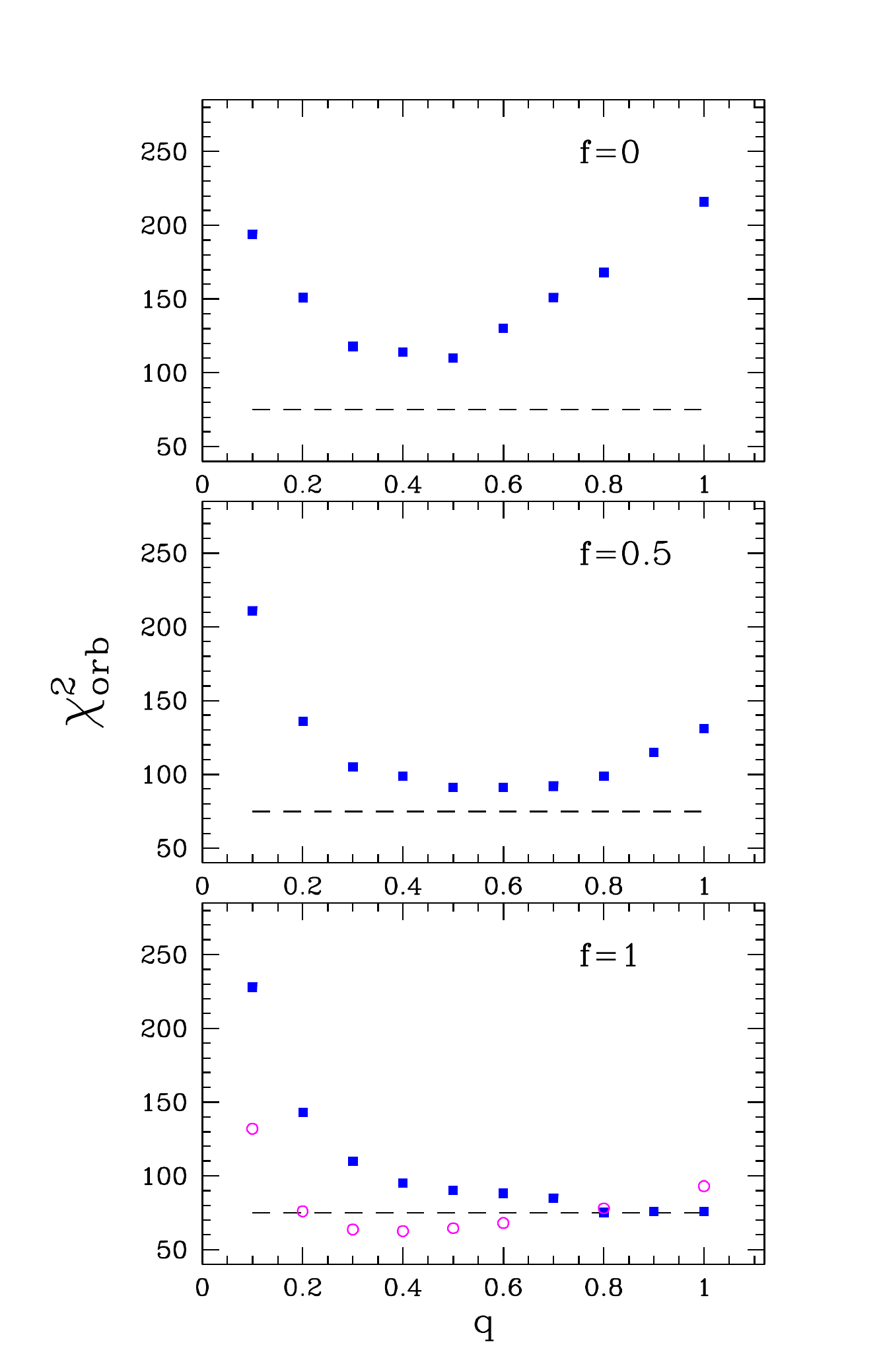}
    \caption{$\chi^2$ residuals for model orbital light curves (18-60 keV) for different mass ratio $q$ and Roche lobe overfilling factor $f=0,0.5,1$. The dashed line indicates the minimal $\chi^2$ for $f=1$. Red open circles shows $\chi^2$ residuals for the model with $f=1$ with taking into account of the 'third light' in the system with a flux of $\simeq 2$ mCrab as derived from the precessional light curve (see Fig. 3).    }
    \label{Fig:deviations}
\end{figure}


As mentioned above, the huge mass transfer rate in SS433 from the donor star ($\dot M_v\sim 10^{-4}\Mdot$) occurring most likely in the thermal time-scale may result in a significant overflow of the inner Roche lobe by the optical star. In this case, it is reasonable to assume that the size of the star is bounded by the outer Roche lobe passing through the L$_2$ point \footnote{Strictly speaking, the star's surface tending to be bounded by an appropriate equipotential surface should strongly deviate from the Roche model in the vicinity of the L$_1$ point where powerful outflow forms a thick 'neck' transforming into the gas stream. However, exact structure of this complicated region is of minor importance for our geometrical modeling of the X-ray eclipse.}. The possibility of mass outflow through the L$_2$ point in SS433 binary system was first suggested by \cite{1993MNRAS.261..241F} from spectral analysis.   

As in the case of exact Roche lobe filling by the donor star described in the previous Section \ref{Susec:exactRoche} \citep{2013MNRAS.436.2004C}, we have used a model of precessing accretion disc with hot corona (see Fig. \ref{Fig:8}) and fixed the binary inclination angle $i=78^o.8$ and accretion disc tilt $\theta=20^o.3$ as derived independently from the analysis of moving emission lines in the SS433 spectrum. The model parameters included $r_d, \omega, a_j, b_j$ and $q$.


Assume that the form of the optical star in SS433 is bounded by one of the equipotentials in the Roche model. The inner Lagrangian and outer Lagrangian surfaces passing through the inner and outer Lagrangian points L$_1$ and L$_2$ correspond to the Roche potential $\Omega_{in}$ and $\Omega_{out}$, respectively. Equipotential surfaces locating between the inner and outer critical surfaces can be labeled by the potential $\Omega$. It is convenient to introduce the degree of contact parameter \citep{1979ApJ...231..502L}:
\begin{equation}
    f=\frac{\Omega-\Omega_{in}}{\Omega_{out}-\Omega_{in}}\,.
\end{equation}
Clearly, the inner and outer Lagrangian surfaces correspond to $f=0$ and $f=1$, respectively, and in between these critical surfaces $0\le f\le 1$.

\begin{figure}
	\includegraphics[width=\columnwidth]{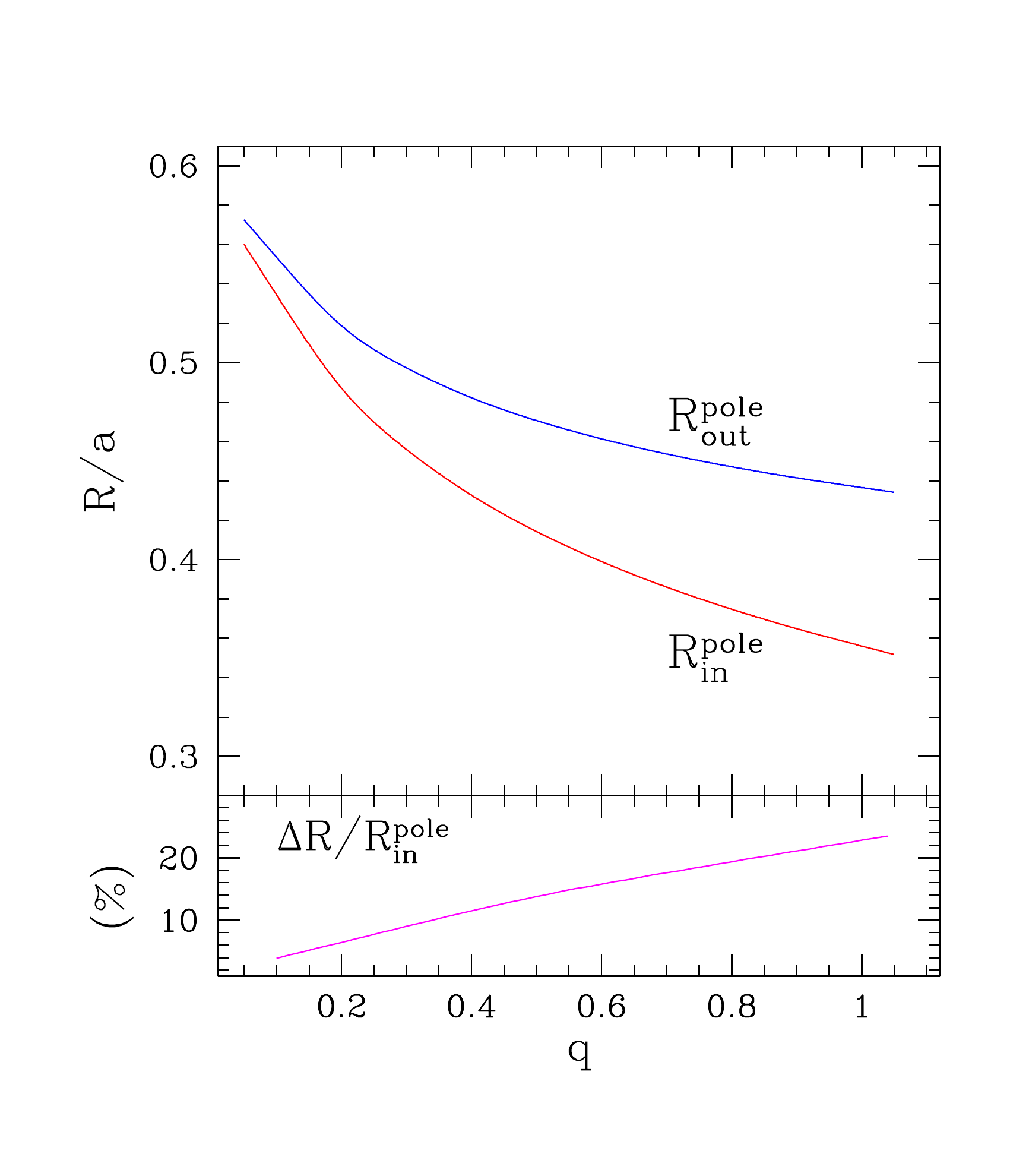}
    \caption{The polar radius of the inner ($R_{in}^{pole}$) and outer ($R_{out}^{pole}$) Lagrangian surfaces and their relative difference $\Delta R$ as a function of the mass ratio $q$.
    }
    \label{Fig:radii}
\end{figure}

\label{Fig:newchi2}
\label{Fig:newchi2-2}
By assuming the size of the star exceeding the inner Lagrangian surface, we performed the modeling of orbital 18-60 keV light curves of SS433 for different mass ratios $q$. In \citep{2013MNRAS.436.2004C} and Section \ref{Susec:exactRoche} above we used the model with $f=0$. Now consider the case with $f=0.5$ and $f=1$. Fig. \ref{Fig:deviations} shows best-fit $\chi^2$ for the model 18-60 keV orbital light curves for different $q$ for parameters providing good model fit to the observed    precessional light curve. Fig. \ref{Fig:deviations}, as the earlier modeling \citep{2013MNRAS.436.2004C}, disfavours low binary mass ratios $q<0.3$.  In the new modeling, $\chi^2$ significantly decreased for $q>0.5$, especially for the optical star filling the outer Lagrangian surface ($f=1$).  This result is understandable because the increased size of the optical star led to the primary eclipse width increase and therefore for $q>0.5$ the model light curves well fit observations. Somewhat unexpected was the fact that for the maximum star size ($f=1$) the $\chi^2$ values decrease for larger $q$, and even for $q=1$ the orbital eclipse width is in agreement with observations (the earlier analysis has implied that for $f=0$ the model primary eclipse is noticeably narrower than the observed one). This effect gets more pronounced for large $q$ while for low $q$ the difference between $\chi^2$ of the residuals in both models is comparatively small. To understand more clearly the results, in Fig. \ref{Fig:radii} we show polar radii of the inner and outer Lagrangian surfaces as a function of the binary mass ratio $q$. 

\begin{figure}
	\includegraphics[width=\columnwidth]
		{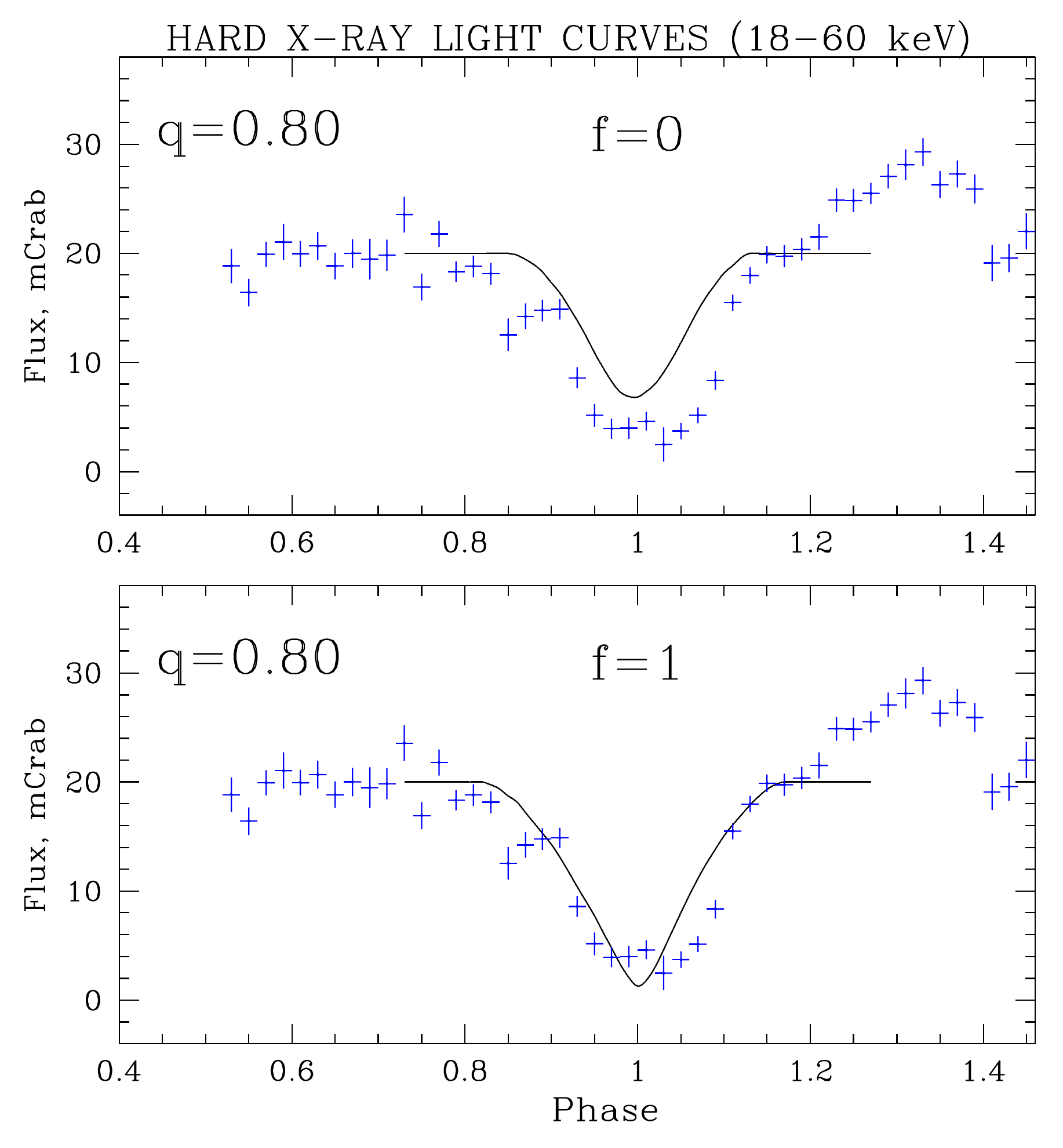}
    \caption{The model eclipse light curve (the solid line) in the case $f=0$ (upper panel) and $f=1$ (bottom panel) for $q=0.8$ with other model parameters corresponding to the best fit of the precessional X-ray light curve. 
    }
    \label{Fig:q08f01}
\end{figure}

With increasing $q$, the size of the inner Lagrangian surface (and correspondingly of the Roche lobe-filling star) decreases, the polar radii $R^{in}_{pole}$ decreases. The polar radius of the outer Lagrangian surface $R_{out}^{pole}$ also decreases with $q$, however, somewhat slower. Thus, $R^{out}_{pole}$ is significantly larger than $R^{in}_{pole}$ (see also Tables in \citealt{1964BAICz..15..165P}). As a result, the strong relative size increase of the star filling the outer Lagrangian surface results in a significant increase in width of the model orbital eclipse  for $q\sim 0.8-1$.  This causes the monotonic decrease of the $\chi^2$ residuals (Fig. \ref{Fig:deviations}, bottom panel) with increasing $q$ for the case $f=1$.     
Fig. \ref{Fig:q08f01} presents the observed and model orbital light curve for $f=0$ and $f=1$ and the same model parameters with $q=0.8$ which best fit the precessional light curve. It is seen that the model eclipse width for $q=0.8$ is larger for the star filling the outer Lagrangian surface ($f=1$) than for the canonical case of the star filling the inner Lagrangian surface (the Roche lobe, $f=0$).   

We should note that the corresponding minimal values of the reduced $\chi^2$ in all cases ($f=0,0.5,1$) are much larger than unity. This means that our purely geometrical model does not fit observational data fully adequately, which can be due to complexity of physical processes in SS433. Nevertheless, the $\chi^2$ minimum can point to the optimal values of the model parameters.  
The minimum of the $\chi^2$ residuals was used to constrain the admissible mass ratio interval. As above, the minimization of the residuals for the orbital light curve for fixed $q$ was performed only for the parameters $r_d, \omega, a_j, b_j$ enabling good fit to the off-eclipse and mid-eclipse precessional light curves of SS433 (see Fig. \ref{Fig:3prec2}).

As in the previous case ($f=0$), for low binary mass ratios $q\sim 0.1-0.2$ the orbital X-ray eclipse can be satisfactorily reproduced only in the model of a thin long X-ray jet ($b_j\gtrsim0.5$), but the model amplitude of the precessional variability turns out to be much smaller than observed. At the same time, by demanding the precessional light curve to be well described for low $q$, the model of a short thin jet predicts the appearance of a noticeable mid-eclipse plateau, which is not observed. As a result, only for $q\gtrsim 0.4$ we are able simultaneously model both precessional and orbital light curve. As seen from Fig. \ref{Fig:deviations}, for $f=1$ the minimum of the residuals is reached for $q\gtrsim 0.8$. 
In this Figure, the dashed line shows the minimum $\chi^2$ for the case $f=1$ where theoretical star fills the outer Lagrangian surface ($\chi^2\simeq 75$).  In other cases ($f=0.5,0$) the minimum $\chi^2$ residuals are significantly larger than 75. The optimal mass ratio $q$ corresponding to the minimum $\chi^2$ residuals are $q\approx 0.4,0.6,0.8$ for $f=0, 0.5, 1$, respectively. 

We stress once again that for small binary mass ratios around $q\simeq 0.1$ the $\chi^2$ residuals increase because in order to well fit the precessional light curves, the model orbital light curve demonstrates a noticeable plateau at the eclipse minimum, which is not observed. At large mass ratios $q\simeq 1$, by requiring a good fit to the precessional light curve, the model orbital light curve for $f=0$ and $f=0.5$  demonstrates a narrower eclipse width than observed. Thus, our joint analysis of the precessional and orbital X-ray light curves of SS433 enables us to choose the optimal binary mass ratio $q=M_x/M_v\ge 0.4-0.8$. 

The precessional off-eclipse X-ray 18-60 keV light curve of SS433 (Fig. 3) suggests that at the 'cross-overs' when the disc is observed edge-on, the flux from the system does not vanishes reaching a minimal value of $\simeq 2$ mCrab. As noted above, this value 2 mCrab should be considered as an upper limit only for the minimum X-ray flux detected by the INTEGRAL telescopes (in this case, 
calibration detector errors could lead to a systematic error). Nevertheless, to check the stability of our mass ratio estimate $q>0.3$, it seems reasonable to consider some generalization of our geometrical eclipsing model of SS433.

It is possible to assume that there is additional  non-eclipsed 'third light' which may be due scattering of hard X-ray emission in the extended disc wind outflow. Therefore, we have performed additional modeling of the 18-60 keV orbital light curve with an account of the 2 mCrab 'third' non-eclipsed component. The results for the case $f=1$ are shown in Fig. \ref{Fig:deviations} by open red circles. Clearly, the residuals decrease in this case, however, as before, the mass ratios $q<0.3$ can be rejected. The optimal mass ratio estimate with an account for the third non-eclipse light is $q\gtrsim 0.3\div 0.4$.  

We emphasize that here we have restricted the mass ratio $q$ by minimizing the residuals $\chi^2_{orb}$ between the observed and model \textit{orbital} light curves of SS433 on the set of model parameters limited by the precessional light curve (see Fig. 12). In our paper \cite{2009MNRAS.397..479C} the limit $q>0.3$ for SS433 was obtained using another method. In those case we have minimized the $\chi^2_{prec}$ residuals between the observed and model \textit{precessional} X-ray light curves of SS433 on the set of model parameters restricted by the orbital eclipsing light curve. Then the best fit of the orbital light curve is obtained for small mass ratios $q\sim 0.1$; however, in that case the height of the 'corona' is quite large. At the cross-overs, the 'corona' is not fully eclipsed by the outer parts of the disc, which unables us to reproduce both the form and amplitude of the precessional light curve. Therefore, mass ratios $q<0.3$ are rejected. Thus, the analysis of the orbital eclipsing light curve only does not enable us to restrict the binary mass ratio $q$. Similarly, the analysis of the precessional light curve only does not enable the binary mass ratio to be restricted.  
Only the \textit{joint} analysis of the orbital and precessional light curves gives the possibility to reliably constrain the binary mass ratio in SS433 $q>0.3$. In
\cite{2009MNRAS.397..479C}, the estimate $q>0.3$ was obtained by assuming that the donor star fills its inner Lagrangian surface ($f=0$). In the case where the optical star fills its outer Lagrangian surface ($f=1$), the limit $q>0.3$ is reinforced. 

To conclude this Section, we note that as the hard X-ray eclipse is wider than that in the standard X-ray band (see Fig. \ref{Fig:9}), 
by solving our inverse problem we always get a wide corona (big half-opening angle $\omega\sim 70^o-80^o$), which can be related to rough description of the form of hot corona filling the supercritical accretion disc funnel. The conical approximation we use in our geometrical model is clearly oversimplified.  

\section{Mass ratio constraints from binary orbital period stability}

The new estimate of the binary mass ratio in SS433 obtained above is in good correspondence with totally independent constraints as inferred from the observed stability of the orbital binary period of this system \citep{2018MNRAS.479.4844C,2019MNRAS.485.2638C}.  

In \cite{2018MNRAS.479.4844C} we have shown for the first time that to provide the observed log-term stability of the orbital period of SS433 in the model of isotropic re-emission from supercritical accretion disc, an account of the possible mass-loss from the outer Lagrangian point L$_2$ always increases the binary mass ratio estimate: $q\gtrsim 0.6$. 
We stress that this method of the binary mass ratio estimate is independent of spectroscopic data of SS433. 

A huge optical luminosity of the supercritical accretion disc in SS433, as well as a powerful disc wind and donor star wind, strongly hamper the reliable determination of the binary mass ratio from spectroscopic measurements (see, e.g., \citealt{2004ASPRv..12....1F,2010A&A...521A..81B,2011A&A...531A.107B,2011MNRAS.417.2401B,2018A&A...619L...4B}). In particular, the selective absorption of the optical star light in the rotating circumbinary shell can misrepresent the spectral classification of this star and distort the radial velocity curve constructed using the absoprtion lines in the optical spectrum. Evidences for the circumbinary shell in SS433 have been found both from radio observations suggesting the presence of an equatorial gas outflow normal to the relativistic jets \citep{2001ApJ...562L..79B,2011A&A...531A.107B} and form infrared and optical spectroscopy revealing the presence of double-peak hydrogen emission lines \citep{1988AJ.....96..242F,2010MNRAS.408....2P,2017ApJ...841...79R}.

The recent spectrophotometric and astrometric observations by the VLTI GRAVITY interferometer \citep{2019A&A...623A..47W} enabled the mapping of the double-peak Brackett Br-$\gamma$ emissions in the infrared spectrum of SS433. The observations suggest that these emissions cannot be formed in an accretion disc around the compact object but are rather produced in a disc-like circumbinary shell demonstrating a super-Keplerian rotation and significant radial expansion. The GRAVITY observations also showed that the broad component of the double-peak emissions in SS433 is formed in the stellar wind outflow from the supercritical accretion disc supporting the composite model of the Brackett emission line profile proposed in \cite{2018MNRAS.479.4844C}. In this model, the double-peak component observed by \cite{2017ApJ...841...79R} is formed in a circumbinary rotating shell and the broad profile is produced in the disc wind outflow. Now this model is confirmed by the direct observations, additionally supporting the possibility of the significant Roche lobe overflow by the optical donor star.

Analysis of optical observations of SS433 \citep{2018ARep...62..747C} that have been performed in the last decades since its discovery does not revealed significant changes in both the binary orbital period and precessional period. This feature is rather surprising because the mass-loss rate from the optical star in this high-mass binary system is among record-high known, $\dot M_v\sim 10^{-4}\Mdot$. The mass outflow rate in the relativistic jets is much smaller, $\dot M_j\sim 10^{-6}\Mdot$. Most of the matter accreting onto the compact object is thought to be expelled in a quasi-spherical wind outflow from a supercritical accretion disc. As noted above, there are also firm observational signatures of outflow in the form of circumbinary disc or jets. All these findings evidence for a significant removal of angular momentum from the binary system, which should be taken into account in the analysis of the orbital stability.

While the exact physical mechanism of angular momentum loss from the system remains unclear, we are able to put independent constraints on the binary mass ratio $q$ from the analysis of the binary orbital period stability over last several decades (see \cite{2018MNRAS.479.4844C,2019MNRAS.485.2638C} for a more detailed discussion). 

\begin{figure}
	\includegraphics[width=\columnwidth]{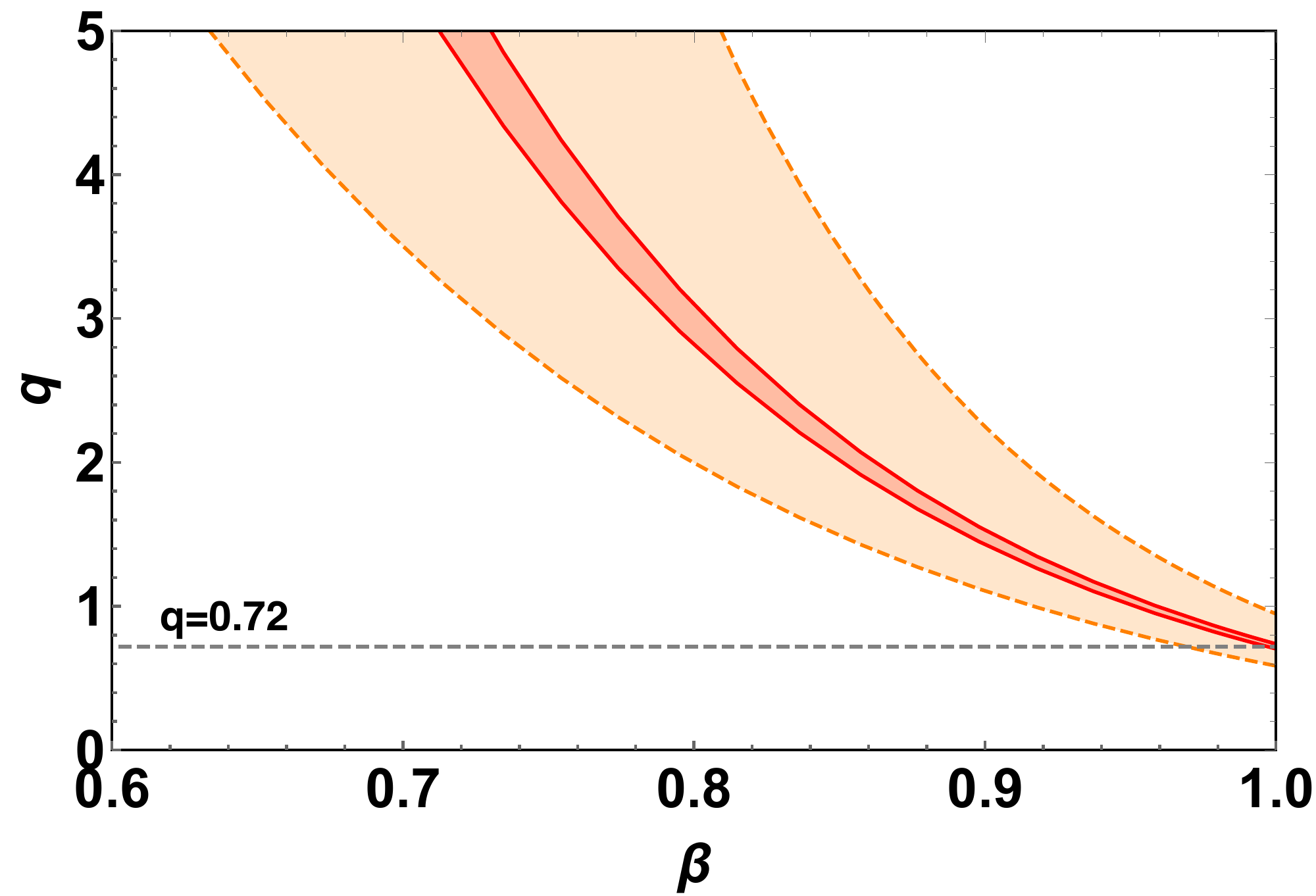}
    \caption{Constraints on the mass ratio of SS433 for different values of the mass-loss rate $\dot M_\mathrm{v}=10^{-4}$ and $10^{-5} M_\odot$ yr$^{-1}$ (intervals restricted by the solid and dashed lines, respectively), for the fiducial value of the dimensionless parameter $K=5$. The dashed horizontal line shows the solution of the quadratic equation $3q^2+2q-3=0$ corresponding to $\dot P_\mathrm{b}=0$ at $\beta=1$ (only Jeans mode mass outflow) independently of the mass-loss rate $\dot M_v$. (From \cite{2019MNRAS.485.2638C}).
    }
    \label{Fig:stability}
\end{figure}

The mass loss from the system occurs in two ways: in the form of isotropic re-emission of the wind outflow from the supercritical accretion disc center \citep{1973A&A....24..337S}(which we will parametrize by the mass-loss rate $\beta \dot M_v$, $\beta\le 1$), and as gas outflow leaving the binary system through  the outer Lagrangian point L$_2$ (which we will parametrize by the mass-loss rate $(1-\beta) \dot M_v$, $\beta\le 1$). In the case of wind outflow from the accretion disc, we can safely assume the specific angular momentum carried away equal to that of the relativistic object, $\omega a_x^2$ ($\omega=2\pi/P_b$ is the orbital angular velocity, $a_x$ the barycentric distance of the relativistic object). The specific angular momentum of matter leaving the system through a circumbinary disc can be directly taken from the GRAVITY VLT observations \citep{2019A&A...623A..47W}, $v_\phi(R_\mathrm{out})R_\mathrm{out}$. Here $R_\mathrm{out}$ is the radius of the circumbinary disc, $v_\phi(R_\mathrm{out})$ is the rotational velocity of the circumbinary disc (super-Keplerian in the case of SS433):
 \[
     \left.\frac{dJ}{dt}\right|_\mathrm{out}=(1-\beta)\dot M_\mathrm{v}v_\phi(R_\mathrm{out})R_\mathrm{out}\,.\nonumber
 \]
 The dominant (major) source of angular momentum in the system is the orbital motion of the binary components, $J_{orb}=(M_xM_v/M)\sqrt{GMa}$ ($M_v$ and $M_x=qM_v$ is the mass of the optical donor star and relativistic object, respectively, $M=M_v+M_x=M_v(1+q)$ is the total mass of the system, $a$ is the orbital separation; the orbit is assumed to be circular). Therefore we can apply the orbital momentum conservation equation:
\[
\frac{dJ_{orb}}{dt}=\beta \frac{\dot M_\mathrm{v}}{M}
\frac{M_\mathrm{v}}{M_\mathrm{x}}
+\left.\frac{dJ}{dt}\right|_{out}\,,
\]
where the first term specifies the angular momentum loss due to isotropic mass re-emission from the supercritical accretion disc characterized by fractional mass-loss $\dot M_{iso}=\beta\dot M_v$,  $0\le\beta\le 1$. 
After straightforward algebra, we arrive at the equation for the fractional change of the binary orbital period \citep{2019MNRAS.485.2638C}:
\[
\frac{\dot P_b}{P_b}=-\frac{\dot M_\mathrm{v}}{M_\mathrm{v}}\frac{3q^2+2q-3\beta -3K(1-\beta)(1+q)^{5/3}}{q(1+q)}\,.
\]
The dimensionless coefficient $K$ specifies the angular momentum loss via circumbinary disc and for the typical values deduced from the GRAVITY VLT observations \citep{2019A&A...623A..47W} is
\begin{eqnarray}
&K=\frac{v_\phi(R_\mathrm{out})R_\mathrm{out}}{(GM_\mathrm{v})^{2/3}}
\bigg(\frac{2\pi}{P_b}\bigg)^{1/3}\approx \nonumber \\
&\approx 5.1 \bigg(\frac{v_\phi(R_\mathrm{out})}{220\,\mbox{km\,s}^{-1}}\bigg)
\bigg(\frac{R_\mathrm{out}}{0.7\mbox{mas}}\bigg)
\bigg(\frac{M_\mathrm{v}}{15\,M_\odot}\bigg)^{-2/3}\,.\nonumber
\end{eqnarray}
From the observed stability of the orbital period of SS433 over $\Delta t\sim 28$ years, $|\dot P_b|\le 3\times \sigma_P/\Delta t$, where $\sigma_P=0^\mathrm{d}.00007$ is the uncertainty in the SS433 orbital period determination  \citep{2011PZ.....31....5G},  and assuming the optical star mass $M_\mathrm{v}=15 M_\odot$ with the mass-loss rate $\dot M_\mathrm{v}=10^{-4} M_\odot$~yr$^{-1}$, in Figure \ref{Fig:stability} we plot the constraints on the binary mass ratio in SS433 $q$ for $K=5$, as a function of the parameter $\beta$. 
Clearly, even a tiny fraction of the total mass loss via the circumbinary shell, $1-\beta\simeq 0.01$, places a mass ratio constraint $q\gtrsim 0.6$ in SS433. This limit is in agreement with our analysis of the hard X-ray orbital and precessional variability of SS433 $q\gtrsim 0.5$
(see \citealt{2013MNRAS.436.2004C} and above in Section \ref{Sec:7modelparam}). Here we stress that allowance for mass-loss through the circumbinary shell only increases the minimum mass ratio $q$ capable of providing the observed stability of the orbital period in SS433.

\begin{figure}
	\includegraphics[width=\columnwidth]{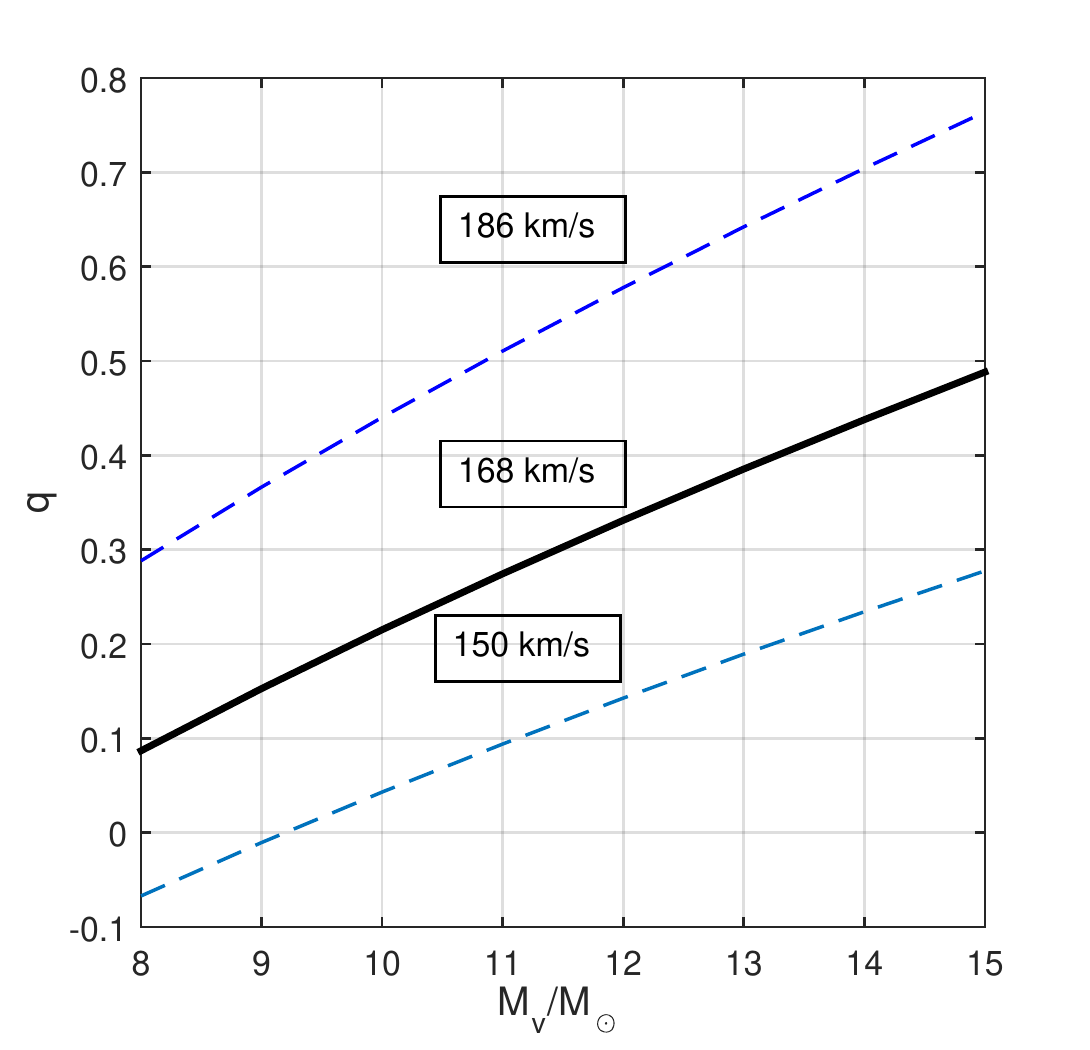}
    \caption{Binary mass ratio $q$ corresponding to the radial orbital velocity of the compact object as required to reproduce observed HeII radial velocity semi-amplitude $K_\mathrm{x}=168\pm 18$\kms. (From \cite{2019MNRAS.485.2638C}). }
    \label{Fig:f16}
\end{figure}

Fig. \ref{Fig:f16} shows the binary mass ratio $q$
as a function of the optical star mass $M_v$  which gives the observed radial velocity semi-amplitude of the compact star estimated from He II emission, $K_\mathrm{x}=168\pm 18$\kms \citep{2004ApJ...615..422H}. The obtained restriction on the binary mass ratio, $q>0.6$, enables us to independently constrain the optical star mass, $M_v\gtrsim 12 \Msol$. This estimate is independent of the distance to the source. Therefore, for $M_v>12 \Msol$ and $q>0.6$ the mass of the compact object in SS433 is $M_x\gtrsim 7 \Msol$, which reinforce the conclusion that the compact object in SS433 is a typical stellar-mass black hole. 
 
\section{Summary of analysis of INTEGRAL observations}

The long time interval spanned by INTEGRAL observations of SS433 (2003-2011) enabled us to obtain new results which are summarized below.

1. For the first time, hard X-ray emission (18-60 keV) was detected from supercritically accreting compact source suggesting the presence of a cone-like funnel in the accretion disc centre through which the accretion-generated hard X-ray emission leaks out from the system \citep{2003A&A...411L.441C}. 

2. A unique feature of SS433 differing this supercritically accreting microquasar from classical X-ray binaries is found: the width of the X-ray source eclipse by the optical star in 18-60 keV band is wider than that in the standard 2-10 keV X-ray band \citep{2005A&A...437..561C}. This evidences for the hard X-ray emission from an extended hot central region of the supercritical accretion disc (a 'corona').  

3. The amplitude of both the orbital eclipse and precessional hard X-ray variability exceeds one order of magnitude suggesting that the 'corona' is almost fully screened by the donor star during the orbital eclipse and is fully eclipsed by the outer parts of the precessing acccretion disc. 

4. For the first time, the hard X-ray spectrum of SS433 was found to keep its shape both during orbital and precessional variability (is described by a power law with the photon index $\Gamma\simeq 3.8$) despite an order-of-magnitude flux change. This strongly suggests that the Comptonized emission from an extended hot quasi-isothermal 'corona' ($kT\simeq 2 0$~keV) and not thermal emission from relativistic jets with rapidly radially decreasing temperature  dominates \citep{2005A&A...437..561C,2009MNRAS.397..479C,2013MNRAS.436.2004C,2009MNRAS.394.1674K}. The evidence of the hot 'corona' in hard X-rays was independently found from \textit{Suzaku} observations of hard X-ray excess in the SS433 spectra \citep{2010PASJ...62..323K}. 

5. The shape of the 18-60 keV orbital light curve of SS433 is found to demonstrate two humps at orbital phases 0.25 and 0.75, especially in 40-60 keV. This feature is likely to be due to the nutation of the accretion disc with a period of 6.290 d. The nutational variability was also supported by the analysis of Swift/BAT light curves of SS433 \citep{2013MNRAS.436.2004C}.
The observed hard X-ray nutational variability of SS433 lends more credence to the production of hard X-ray emission of SS433 in an extended hot corona. 

6. For the first time, the INTEGRAL observations of SS433 revealed the presence of the secondary maximum in the precessional light curves, visible both in the broad-band 18-60 keV and more narrow 18-40 and 40--60 keV bands.  This finding directly suggests that the hot extended corona is visible in moments of both maximum disc opening ($\psi_{prec}=0$) and between two cross-overs ($\psi_{prec}=0.34, 0.66$) when the disc is turned by the 'rear' side to the observer. These fine details of the precessional variability support our model of the supercritical accretion disc with extended hot central 'corona'. 

7. INTEGRAL observations discovered a significant (up to a factor of two) variability of the width of X-ray eclipse of the accretion disc by the donor star \citep{2009MNRAS.397..479C} implying that the eclipse is not a purely geometric effect. The complex eclipse form evidences for the important role of absorption effects due to gas streams forming the precessing accretion disc and also appearing in the wind collision region between the donor star and the disc. The upper envelope of the eclipse ingress is found to be the most appropriate for geometrical modeling of the hard X-ray light curve.

8. For the first time, a joint interpretation of the orbital and precessional hard X-ray light curves is performed in the model of precessing accretion disc with hot extended corona assuming the donor star's filling of both inner (passing through the L$_1$ point) and outer (passing through the L$_2$ point) Lagrangian surface. This analysis made it possible to conclude that low values of the binary component mass ratio $q=M_x/M_v<0.3$  can be reliably rejected by 18-60 keV INTEGRAL observations \citep{2013MNRAS.436.2004C}. The possibility of the donor star to stably overfill its Roche lobe shifts the lower limit of admissible binary mass ratio to $q_{min}\simeq 0.5$. Thus, our modelling of the INTEGRAL orbital and precessional light curves suggests $q\gtrsim 0.3\div 0.5$, which can be considered as conservative mass ratio estimate. 

9. Our recent analysis of the observed stability of the binary orbital period over $\sim 30$ yrs taking into account of the angular momentum loss from the system due to isotropic re-emission from the central parts of the supercritical accretion disc and mass loss through the L$_2$ point \citep{2018MNRAS.479.4844C,2019MNRAS.485.2638C} suggests $q\gtrsim 0.6$. Here we have used the results of the VLT GRAVITY observations \citep{2019A&A...623A..47W} discovered a circumbinary shell around SS43 rotating with a super-Keplerian velocity. 

10. We emphasize the agreement between the two independent estimates of the binary mass ratio  $q\gtrsim 0.3\div 0.5$ and $q\gtrsim 0.6$ discussed above. Assuming the donor star mass in SS433 to be $M_v=8\div 15\,\Msol$, our estimate $q>0.6$ leads to the mass of the relativistic object in SS433 (irrespective of the spectral analysis of the system) $M_x=qM_v>(5\div 9)\,\Msol$, which provides solid grounds to consider the relativistic object in SS433 as a black hole with mass close to the mean mass of black hole candidates in Galactic X-ray binaries ($\sim 8\,\Msol$). 

11. The relatively high binary mass ratio in SS433 obtained from our analysis $q>0.6$ is in agreement with conclusions by \cite{2017MNRAS.471.4256V} about the absence of the common envelope in this system thus confirming the evolutionary status of SS433 as a massive semi-detached X-ray binary at late stages of the secondary mass exchange with stable mass transfer rate from the donor star onto the relativistic object, most likely a black hole. Note that the high mass of the compact object in SS433 was recently inferred by \cite{2018A&A...619L...4B} from different considerations. 

\cite{2017MNRAS.471.4256V} note that if the initial mass of the donor star in SS433 exceeds $\sim 11-12\,\Msol$, the evolution of SS433 ends up with the formation of a (neutron star+black hole) binary system. As our analysis suggests the optical star mass above $\sim 12 \,\Msol$, the possibility of the ultimate formation of a NS+BH or even BH+BH binary from SS433 is reinforced. If compact enough, such a binary would merge due to gravitational radiation losses to become a binary coalescence gravitational wave source.

\section{Conclusions and outlook}

Based on the INTEGRAL observations, we have considered in detail the properties of SS433 as a massive X-ray binary passing through the final stages of the secondary mass exchange between the components. By analyzing hard X-ray INTEGRAL observations of SS433 and using the fact of the observed constancy of the orbital binary period of SS433 with an account of recent IR observations of SS433 by the VLT GRAVITY interferometer, we have obtained reliable constraints on the binary mass ratio $q=M_x/M_v\gtrsim 0.6$ suggesting the black hole nature of the relativistic companion in this binary system. Here we did not considered many other intriguing properties of SS433 yet to be explained. The unsolved issues of SS433 include, in particular, the following.

1. The mechanism of the accretion disc precession with a period of $\sim 162.3$ d. Tilted accretion discs precessing with a period an order of magnitude longer than the orbital one are known in different classical X-ray binaries, including Her X-1 ($P_{prec}\simeq 35$ d \citep{1972ApJ...174L.143T}), Cyg X-1 ($P_{prec}\simeq 147$ d \citep{2011MNRAS.412.1985Z}), LMC X-4 ($P_{prec}\simeq 30.4$ d \citep{1981ApJ...246L..21L}), etc. 

An orbital tilt of the disc can be due to asymmetric supernova explosion leading to the formation of the relativistic component that can turn the orbital plane relative to the rotational axis of the optical star \cite{1974ApJ...187..575R,1981SvAL....7..401C}. The tilted disc should precess because the optical star's spin axis precesses (the 'slaved' precession, \citealt{1973SvA....16..756S}). The disc will also undergo forced tidal precession \citep{1973NPhS..246...87K}. The precessing disc in SS433 was discussed by \cite{2004ASPRv..12....1F} (see also references in that review). The stability of kinematic model parameters of SS433 observed over 40 years, as well as small variability in the jet outflow velocity, favours the slaved precession model \citep{2018ARep...62..747C}. The low-amplitude orbital periodicity in the jet velocity \citep{2018ARep...62..747C} may imply that the orbit of SS433 is slightly eccentric \citep{2007A&A...474..903B}, which is in agreement with the slaved precessing accretion disc following the precession of the optical star spin axis \citep{1980A&A....81L...7V}. The point is that in the slaved accretion disc model for SS433 the optical star spin axis is tilted to the orbital plane by $\sim 20^o$ implying asynchronous rotation of the star with the orbital period. However, tidal synchronization model \citep{1977A&A....57..383Z,1989A&A...220..112Z} suggests that the tidal synchronization of the binary components spins occurs earlier than the orbital circularization. Therefore, if the optical star in SS433 is in asynchronous rotation with the orbital period, the binary orbit could be slightly eccentric, which can be the case for SS433. The optical light curves of SS433 are subject to significant physical variability from period to period which makes it difficult to detect a low orbital eccentricity $e<0.1$ quite admissible by the optical observations.

2. The collimation mechanism of relativistic jets up to the observed velocity $\sim 80000 \,\kms$ is another puzzle. The jet velocity remains on average stable over $\sim 40$ years \citep{2018ARep...62..747C}. SS433 turned out to be the first representative of a new class of objects -- Galactic microquasars, -- X-ray binaries exhibiting collimated  relativistic jets with bulk velocities up to $0.9 c$ (massive X-ray binary Cyg X-1, low-mass X-ray binaries GRS1915+105, GROJ1655-40, etc.). The salient feature of SS433 jets is their precession and quasi-stationarity. The jet precession is closely related to that of the accretion disc (observations suggest that jets are perpendicular to the disc plane and follow its preccessional motion). The constant jet velocity can be due to the line-locking effect \citep{1979ApJ...230L..41M,1982IAUS...97..209S}.

3. Recently, the detection of GeV gamma-ray emission from the SS433 region  was reported from the analysis of ten-year's \textit{Fermi-LAT} observations \citep{2019ApJ...872...25X}. This emission is likely to be formed in the jet interaction region with the ambient W50 nebula. This finding was independently confirmed by the analysis of \textit{Fermi-LAT} 200-500 MeV data \citep{2019MNRAS.485.2970R}, who also securily detected the precessional variability of the gamma-ray emission. At very-high-energy range ($\sim 20$ TeV), SS433 was detected by High Altitude Water Cherenkov (HAWC) observatory \citep{2018Natur.562...82A}.
VHE from the jet/W50 interaction region was detected by MAGIC and H.E.S.S. Cherenkov telescopes \citep{2018A&A...612A..14M}, but no TEV photons were detected from the central binary source. These VHE observations enabled new insights into the physics of particle acceleration and magnetic field in the jet/W50 interaction region. 
However, the rich physics and observations of SS433 jets from radio to VHE range is beyond the scope of the present review (see \citep{2004ASPRv..12....1F} for more discussion of jets and early references). 

4. In recent years, radiation hydro simulations of supercritical accretion discs have been performed (see, e.g., \citep{2005ApJ...628..368O,2009MNRAS.398.1668O,2011ApJ...736....2O}). 
The similarity of SS433 with the ultraluminous X-ray sources (ULXs) has long been stressed \citep{2001IAUS..205..268F,2001ApJ...552L.109K,2004ASPRv..12....1F} and the supercritically accreting HMXBs with stellar-mass black holes remain the possible model for ULX. 
Presently, strongly shifted spectral lines suggesting the presence of relativistic jets were reported in two ultra-luminous X-ray sources, M81ULX-1 and NGC300 ULS-1 ($v/c\simeq 0.17-0.24$) \citep{2015Natur.528..108L,2018MNRAS.479.3978K}. \cite{2013Natur.504..260D} reported the discovery of baryonic jets from putative black hole in X-ray binary 4U17630-70 that can be an SS433 analog. Recently, the structure of emission regions around relativistic jets in SS433 was probed by the GRAVITY VLTI interferometer \citep{2018arXiv181112564W}. 

Finally, we should mention recent claims \citep{2017ASPC..510..466G} on the possible rapid binary evolution in SS433 appearing as systematic changes of its photometric variability. Here we stress again, however, that our analysis of parameters of the kinematic model of SS433 inferred from spectroscopy of moving emission lines (including precessional, nutational and orbital periods, the module of the jet outflow velocity and the jet inclination angle to the orbital axis) suggest that they remain constant on average over $\sim 40$ years of observations \citep{2018ARep...62..747C}. Therefore, further spectral and photometric multiwavelength observations of SS433 are important and strongly encouraged. X-ray observations of SS433 and surrounding nebula W50 are already scheduled in the forthcoming eROSITA mission. 

\section*{Acknowledgements}

The work of ACh and EA (the analysis of the X-ray variability) is supported by the RSF grant 17-12-01241. The work of KP and AB  is supported by the Program of development of M.V. Lomonosov Moscow State University (Leading Scientific School 'Physics of stars, relativistic objects and galaxies'). KP also acknowledges the support from the RFBR grant No. 19-02-00790.  
 
\section*{References}



\bibliographystyle{elsarticle-harv} 
\bibliography{SS433.bib}

\end{document}